\documentclass{emulateapj}
\usepackage{subfigure}

\def\icnoblXicblLogM{0.8\%}

\def\numicnoblLogM{42}

\def\icblXicnoblLogM{0.8\%}

\def\icblXiiLogM{0.4\%}

\def\icblXiibLogM{13\%}

\def\icblXibLogM{0.1\%}

\def\numicblLogM{16}
\def\iiXicnoblLogM{19\%}

\def\iiXibicnoblLogM{0.5\%}

\def\iiXibLogM{0.7\%}

\def\iibXicnoblLogM{47\%}

\def\iibXiiLogM{66\%}

\def\iibXibLogM{10\%}

\def\numiibLogM{22}

\def\numibLogM{36}

\def\icnoblXicblLogMtar{64\%}

\def\numicnoblLogMtar{37}

\def\numicblLogMtar{7}

\def\iibXibLogMtar{14\%}

\def\numiibLogMtar{19}

\def\numibLogMtar{26}

\def\icnoblXicblLogMuntar{63\%}

\def\numicnoblLogMuntar{5}

\def\numicblLogMuntar{9}

\def\iibXibLogMuntar{9.5\%}

\def\numiibLogMuntar{4}

\def\numibLogMuntar{9}

\def\icnoblXicblLogMpro{0.005\%}

\def\numicnoblLogMpro{30}

\def\numicblLogMpro{11}

\def\iibXibLogMpro{7.9\%}

\def\numiibLogMpro{14}

\def\numibLogMpro{25}

\def\icnoblXicblsnrallhighmass{16\%}

\def\numicnoblsnrallhighmass{36}

\def\icblXicnoblsnrallhighmass{16\%}

\def\icblXiisnrallhighmass{95\%}

\def\icblXiibsnrallhighmass{93\%}

\def\icblXibsnrallhighmass{26\%}

\def\numicblsnrallhighmass{7}
\def\iiXicnoblsnrallhighmass{2.0\%}

\def\iiXibsnrallhighmass{2.5\%}

\def\iibXicnoblsnrallhighmass{0.4\%}

\def\iibXiisnrallhighmass{12\%}

\def\iibXibsnrallhighmass{0.7\%}

\def\numiibsnrallhighmass{15}

\def\ibXiibsnrallhighmass{0.7\%}

\def\numibsnrallhighmass{33}

\def\icnoblXicblsnrallhighmasstar{1.6\%}

\def\numicnoblsnrallhighmasstar{35}

\def\numicblsnrallhighmasstar{5}

\def\iibXibsnrallhighmasstar{1.7\%}

\def\numiibsnrallhighmasstar{16}

\def\numibsnrallhighmasstar{26}

\def\icnoblXicblsnrallhighmassuntar{9.7\%}

\def\numicnoblsnrallhighmassuntar{2}

\def\numicblsnrallhighmassuntar{2}

\def\numiibsnrallhighmassuntar{0}

\def\numibsnrallhighmassuntar{6}

\def\icnoblXicblsnrallhighmasspro{20\%}

\def\numicnoblsnrallhighmasspro{25}

\def\numicblsnrallhighmasspro{2}

\def\iibXibsnrallhighmasspro{16\%}

\def\numiibsnrallhighmasspro{9}

\def\numibsnrallhighmasspro{22}

\def\icnoblXiisnrall{0.5\%}

\def\iiXiinsnrall{12\%}

\def\ibXicnoblsnrall{67\%}

\def\icblXicnoblclosethreekpctohfour{2.3\%}

\def\icblXibclosethreekpctohfour{1.6\%}

\def\iiXicnoblclosethreekpctohfour{9.4\%}

\def\iiXibclosethreekpctohfour{60\%}

\def\icblXicnoblclosethreekpcppohfour{0.02\%}

\def\icblXibclosethreekpcppohfour{0.1\%}

\def\iiXicnoblclosethreekpcppohfour{3.0\%}

\def\iiXibclosethreekpcppohfour{26\%}

\def\icnoblXicblcloseppohfour{0.4\%}
\def\icnoblXiicloseppohfour{5.2\%}

\def\numicnoblcloseppohfour{26}

\def\iinXiicloseppohfour{53\%}

\def\numicblcloseppohfour{8}

\def\iibXibcloseppohfour{0.9\%}

\def\numiibcloseppohfour{13}

\def\numibcloseppohfour{16}

\def\icnoblXicblcloseppohfourtar{15\%}

\def\numicnoblcloseppohfourtar{23}

\def\numicblcloseppohfourtar{5}

\def\iibXibcloseppohfourtar{1.8\%}

\def\numiibcloseppohfourtar{13}

\def\numibcloseppohfourtar{11}

\def\icnoblXicblcloseppohfouruntar{2.1\%}

\def\numicnoblcloseppohfouruntar{4}

\def\icblXicnoblcloseppohfouruntar{2.1\%}

\def\numicblcloseppohfouruntar{3}

\def\iibXibcloseppohfouruntar{15\%}

\def\numiibcloseppohfouruntar{1}

\def\numibcloseppohfouruntar{4}

\def\icnoblXicblcloseppohfourpro{0.07\%}

\def\numicnoblcloseppohfourpro{19}

\def\numicblcloseppohfourpro{4}

\def\iibXibcloseppohfourpro{19\%}

\def\numiibcloseppohfourpro{6}

\def\numibcloseppohfourpro{7}

\def\icnoblXicblclosetohfour{5.4\%}
\def\icnoblXiiclosetohfour{1.4\%}

\def\numicnoblclosetohfour{26}

\def\iinXiiclosetohfour{35\%}

\def\numicblclosetohfour{8}

\def\iibXibclosetohfour{8.5\%}

\def\numiibclosetohfour{13}

\def\ibXiiclosetohfour{15\%}

\def\numibclosetohfour{16}

\def\icnoblXicblclosetohfourtar{47\%}

\def\numicnoblclosetohfourtar{23}

\def\numicblclosetohfourtar{5}

\def\iibXibclosetohfourtar{13\%}

\def\numiibclosetohfourtar{13}

\def\numibclosetohfourtar{11}

\def\icnoblXicblclosetohfouruntar{2.1\%}

\def\numicnoblclosetohfouruntar{4}

\def\icblXicnoblclosetohfouruntar{2.1\%}

\def\numicblclosetohfouruntar{3}

\def\iibXibclosetohfouruntar{15\%}

\def\numiibclosetohfouruntar{1}

\def\numibclosetohfouruntar{4}

\def\icnoblXicblclosetohfourpro{0.2\%}

\def\numicnoblclosetohfourpro{19}

\def\numicblclosetohfourpro{4}

\def\iibXibclosetohfourpro{46\%}

\def\numiibclosetohfourpro{6}

\def\numibclosetohfourpro{7}

\def\icnoblXiiSSFRhighsn{3.4\%}

\def\ibicXiiSSFRhighsn{0.3\%}

\def\icblXiiSSFRhighsn{3.5\%}

\def\ibXicnoblSSFRhighsn{26\%}

\def\ibXiinobSSFRhighsn{17\%}

\def\ibicXiiSSFRnuc{10\%}

\def\icnoblXicbluzkc{0.09\%}

\def\icnoblXiibuzkc{0.2\%}

\def\numicnobluzkc{42}

\def\icblXicnobluzkc{0.09\%}

\def\icblXiiuzkc{0.04\%}

\def\icblXibuzkc{0.04\%}

\def\numicbluzkc{16}

\def\iiXiibuzkc{0.1\%}

\def\iibXicnobluzkc{0.2\%}

\def\iibXiiuzkc{0.1\%}

\def\iibXibuzkc{0.03\%}

\def\numiibuzkc{22}

\def\ibXiibuzkc{0.03\%}

\def\numibuzkc{36}

\def\icnoblXicbluzkctar{10\%}

\def\numicnobluzkctar{37}

\def\numicbluzkctar{7}

\def\iibXibuzkctar{0.8\%}

\def\numiibuzkctar{19}

\def\numibuzkctar{26}

\def\icnoblXicbluzkcuntar{4.9\%}

\def\numicnobluzkcuntar{5}

\def\numicbluzkcuntar{9}

\def\iibXibuzkcuntar{0.9\%}

\def\numiibuzkcuntar{4}

\def\numibuzkcuntar{9}

\def\icnoblXicbluzkcpro{0.02\%}

\def\numicnobluzkcpro{30}

\def\numicbluzkcpro{11}

\def\iibXibuzkcpro{1.9\%}

\def\numiibuzkcpro{14}

\def\numibuzkcpro{25}

\def\icnoblXicblukc{3.8\%}
\def\icnoblXiiukc{0.6\%}

\def\iiXicnoblukc{0.6\%}

\def\ibXicnoblukc{55\%}

\def\ibXiiukc{0.6\%}

\def\iiXibicnoblAv{2.1\%}

\def\iibXibicnoblAv{5.1\%}

\def\icnoblXiinztar{1.1\%}

\makeatletter

\newcommand\ionpat[2]{#1$\;${\scshape{#2}}}%

\def\fiboffsetall{$0.45 \times r_{half-light}$}
\def\sniiagn{20$\pm$3\% (36/156)}

\def\snibcagn{9$\pm$4\% (5/55)}

\def\droutibic{45\%}
\def\droutii{2.4\%}
\def\bhnumber{519}

\def\commentslongone{Discov. Method notes whether the SN was discovered by a targeted (T) or a galaxy-impartial (I) search program.
Discov. Pro/Am classifies the discoverers as professional (P) or amateur (A). A superscript number next to a type provides the origin of any 
spectroscopic reclassification (1: Modjaz et al. 2012, in prep.; 2: \citealt{sandersuntargeted}; 3: this work). Discoverer is taken from the IAU classification\footnote{http://www.cfa.harvard.edu/iau/lists/Supernovae.html}.}

\def\commentslongtwo{Offset Norm. is the deprojected offset normalized by the host galaxy half-light radius. Fib. Off. (AGN) is the normalized 
offset of the SDSS spectroscopic fiber with offset most similar to the SN offset that has sufficient S/N to estimate dust extinction and SSFR. 
A superscript number next to a type provides the origin of any spectroscopic reclassification (1: Modjaz et al. 2012, in prep.; 2: \citealt{sandersuntargeted}; 3: this work). 
Fib. Off. (Metal) is the normalized offset of the SDSS spectroscopic fiber with offset most similar to the SN offset that 
is classified as star-forming which enables an oxygen abundance estimate. }

\def\hubbleconstant{73 km s$^{-1}$ Mpc$^{-1}$}

\shorttitle{Core-Collapse SN Environments}

\shortauthors{Kelly \& Kirshner}
\begin{document}

\title{Core-Collapse Supernovae and Host Galaxy Stellar Populations}


\email{pkelly3@stanford.edu}
\author{Patrick L. Kelly\altaffilmark{1,2}}
\author{Robert P. Kirshner\altaffilmark{3}}

\altaffiltext{1}{Kavli Institute for Particle Astrophysics and Cosmology, Stanford University, 382 Via Pueblo Mall, Stanford, CA 94305}
\altaffiltext{2}{SLAC National Accelerator Laboratory, 2575 Sand Hill Rd, Menlo Park, CA 94025}
\altaffiltext{3}{Harvard-Smithsonian Center for Astrophysics, 60 Garden St., Cambridge, MA 02138}

 
\begin{abstract}
We have used images and spectra of the Sloan Digital Sky Survey to examine the host galaxies of 
\bhnumber~nearby supernovae. The colors at the sites of the explosions, as well as chemical abundances, and
specific star formation rates of the host galaxies provide circumstantial evidence on the origin of each
supernova type. We examine separately SN II, SN IIn, SN IIb, SN Ib, SN Ic, and SN Ic with broad lines (SN Ic-BL).
For host galaxies that have multiple spectroscopic fibers, we select the fiber with host radial offset most similar to that of the SN. 
Type Ic SN explode at small host offsets, and their hosts have exceptionally strongly star-forming,
metal-rich, and dusty stellar populations near their centers. The SN Ic-BL and SN IIb explode in exceptionally
blue locations, and, in our sample, we find that the host spectra for SN Ic-BL show lower average oxygen abundances than those for SN Ic.
SN IIb host fiber spectra are also more metal-poor than those for SN Ib, although a significant difference exists for only one
of two strong-line diagnostics. SN Ic-BL host galaxy emission lines show strong central specific star formation rates. 
In contrast, we find no strong evidence for different environments for SN IIn compared to the
sites of SN II. Because our supernova sample is constructed from a variety of sources, there is always a risk that sampling methods can 
produce misleading results.  We have separated the supernovae discovered by targeted surveys from those discovered 
by galaxy-impartial searches to examine these questions and show that our results do not 
depend sensitively on the discovery technique.
\end{abstract}

\maketitle

\section{Introduction}


The only supernovae (SN) found in passive, elliptical galaxies are Type Ia (\citealt{vanden91}; \citealt{vandenbergh05}). Finding these events in galaxies without ongoing star formation is strong evidence that long-lived (or relatively long-lived) progenitors contribute to the observed SN Ia population. SN of other spectroscopic types have 
been discovered only in star-forming galaxies: that is why we think these SN types are explosions of massive, short-lived stars. 

Our aim here is to use more detailed information on the hosts to help sort out the origin of the varieties of core-collapse events.  
Host galaxy measurements have started to identify patterns among the environments 
of the many spectroscopic types of core-collapse supernovae (e.g., \citealt{vd96}; \citealt{mod08}; \citealt{kel08}). Here, we construct a 
nearby sample of supernova hosts where ground-based images provide useful spatial resolution: for the median redshift in our sample 
($z$ $\approx$ 0.02), one arcsecond corresponds to 400 parsecs assuming $H_{o}$=\hubbleconstant. 
We use images from SDSS Data Release 8 (DR 8) 
to measure the color at the supernova sites and to estimate 
the hosts' stellar masses, and Sloan spectra to 
determine the hosts' oxygen abundances, specific star formation rate (SFR), and the interstellar reddening. Although these are blunt tools for determining how star formation, stellar evolution, mass loss, 
and progenitor chemistry produce the diversity of core-collapse phenomena, circumstantial evidence can provide 
useful clues to these complex processes.

The primary SN spectroscopic classes are organized around evidence of hydrogen and helium features (see \citealt{fil97} for a review). 
Young, massive stars with an intact hydrogen envelope at the time of their explosion yield hydrogen-rich spectra, the Type II class.
When massive SN progenitors lose their hydrogen-rich shells, the explosion can produce a variety of spectroscopic outcomes.
SN Ib have a hydrogen-deficient spectrum that shows helium features, while SN Ic do not show either hydrogen or helium lines. 
The chameleon SN IIb class shows the hydrogen lines of a SN II at first, but then shows
helium lines, suggesting there is only a low-mass layer of hydrogen on the surface. Line widths are also important. 
The spectra of SN Ic sometimes show very broad lines, suggesting 
expansion of the surface at 0.1$c$: these are the broad-lined SN Ic (SN Ic-BL). Conversely, SN II are
sometimes seen with exceptionally narrow lines. These are the SN IIn, which result from interaction between the 
ejecta and circumstellar matter. 
SN Ia are a distinct class whose spectra are characterized by the absence of hydrogen and 
the presence of a broad absorption feature at 6150\AA~that 
is attributed to Si. Unlike all the others, they are attributed 
to thermonuclear explosions in white dwarfs.

Spectra of the SN are reported by their discoverers or, in many cases, by independent teams. The 
CfA Supernova program aims to obtain spectra of all the SN north of -20$^{\circ}$ and brighter than 18th 
mag (e.g., \citealt{ma08}; \citealt{blondin12}), following up discoveries made by amateurs and by programs like the Lick Observatory Supernova Search (LOSS; \citealt{filippenko01}; \citealt{fili03conf}).
Programs with the MMT, Magellan, and Gemini pursue fainter SN discoveries from the wide-field PAN-STARRS survey \citep{kaiser10}.
The Palomar Transient Factory (PTF) \citep{law09}, a galaxy-impartial search with the 1.2m Oschin Telescope begun in 2009, has discovered and spectroscopically classified more than a thousand SN. 
Supernova classifications have become more refined over time and the brief reports in IAU Circulars or in catalogs may need to be revisited as new varieties are defined.  For this reason, it will be valuable to archive spectra for future analysis, not just present the classification in an IAU Circular.  The CfA spectra, and collections of other spectra organized by the University of Oklahoma and the Weizmann Institute, are currently available online\footnote{http://www.cfa.harvard.edu/supernova/SNarchive.html}$^{,}$\footnote{http://suspect.nhn.ou.edu/$\sim$suspect/}$^{,}$\footnote{http://www.weizmann.ac.il/astrophysics/wiserep/}. 

With the uniform spectroscopy and $u'g'r'i'z$ imaging of the SDSS, we study the environments of the most populous SN types, identifying a series of strong patterns for stripped-envelope SN types.
Section \ref{sec:data} describes the data. 
Section \ref{sec:sample} details the construction of the SN sample, classification of the SN, and explains how we categorize SN surveys.
We distinguish between the methods of SN discovery and test hypotheses using only targeted or only galaxy-impartial 
SN.  
In Section~\ref{sec:systematics}, we use the redshift distributions of galaxy-impartial SN discoveries to test for detection-related systematic 
effects. 
The measurement of host galaxy photometry and spectroscopic oxygen abundances is described in Section \ref{sec:methods},
our statistical method is described in Section \ref{sec:statmeth}, and Section \ref{sec:results} presents the results of our analysis. 
We compare the relative rates of stripped-envelope SN to Type II with increasing host metallicity to model predictions in Section \ref{sec:rates}.
Section~\ref{sec:tests} presents a discussion of potential systematic effects. 
Finally, we discuss our results in Section \ref{sec:discussion} and present conclusions in Section \ref{sec:conclusion}.



\begin{deluxetable}{lcccccc}
\tablecaption{Galaxy-Impartial Sample Construction}
\tablecolumns{6}
\tablehead{\colhead{Criterion}&\colhead{II}&\colhead{IIn}&\colhead{IIb}&\colhead{Ib}&\colhead{Ic}&\colhead{Ic-BL}}
\startdata
CC/Asiago+PTF/$z<0.08$&116&10&10&11&23&9\\
Discovered 1990-Present&116&10&10&11&23&9\\
Confident Spec. Type&113&10&9&11&21&9\\
Not Ca-rich&113&10&9&11&21&9\\
SN Position&113&10&9&11&21&9\\
DR8 Imaging Footprint&97&9&9&10&17&9\\
Sufficient Coverage&94&9&8&10&17&9\\
No Host Detected&91&7&6&9&17&9\\
\hline
\multicolumn{6}{c}{Host Photometry Sample}\\
No Bright Star&90&7&6&9&16&9\\
No SN Contamination&79&6&5&9&15&9\\
\hline
\multicolumn{6}{c}{Host Spectroscopy Sample}\\
Host Fiber&41&1&1&5&8&3\\
No AGN Contam.&34&1&1&4&8&3\\
\hspace{3mm}Nuclear Fiber&25&1&1&3&7&2\\
\hspace{3mm}Offset within 3 kpc&20&1&1&3&6&3
\enddata
\tablecomments{See Table~\ref{tab:selectionTargeted} description.}
\label{tab:selectionImpartial}
\end{deluxetable}

\begin{deluxetable}{lcccccc}
\tablecaption{Targeted Sample Construction}
\tablecolumns{6}
\tablehead{\colhead{Criterion}&\colhead{II}&\colhead{IIn}&\colhead{IIb}&\colhead{Ib}&\colhead{Ic}&\colhead{Ic-BL}}
\startdata
CC/Asiago/$z<0.023$&577&71&46&70&118&11\\
Discovered 1990-Present&486&65&45&61&113&11\\
Confident Spec. Type&467&59&40&58&105&11\\
Not Ca-rich&467&59&40&57&105&11\\
SN Position&455&58&40&57&102&11\\
DR8 Imaging Footprint&247&36&22&34&57&9\\
Sufficient Coverage&234&34&19&31&55&8\\
No Host Detected&234&34&19&31&55&8\\
\hline
\multicolumn{6}{c}{Host Photometry Sample}\\
No Bright Star&230&34&19&30&55&8\\
No SN Contamination&204&29&17&27&43&7\\
\hspace{3mm}Amateur&90&14&9&10&16&5\\
\hline
\multicolumn{6}{c}{Host Spectroscopy Sample}\\
Host Fiber&115&18&15&15&27&5\\
No AGN Contam.&91&13&12&12&26&5\\
\hspace{3mm}Nuclear Fiber&42&8&7&6&13&3\\
\hspace{3mm}Offset within 3 kpc&59&6&4&7&17&3\\
\hspace{3mm}Amateur&46&3&8&8&11&4
\enddata
\tablecomments{SN remaining of each spectroscopic type after applying inclusion criteria. Indented rows are subsets of the last unindented row.  (1) SN collected in the Asiago Catalog updated through 2010 November 7 with $z<0.023$ for targeted discoveries (and, for Table~\ref{tab:selectionImpartial}, $z < 0.08$ Asiago galaxy-impartial discoveries combined with the 72 Palomar Transient Factory (PTF) core-collapse SN discoveries from March 2009 through March 2010 \citep{arcavi10}) and not classified as Type Ia; (2) SN discovered during period 1990-present to eliminate most discoveries made using photographic plates; (3) Asiago catalog or PTF SN classification not accompanied by (`?'; ambiguous identification) or (`:'; type inferred from light curve not spectrum); 
(4) calcium-rich SN 2000ds, SN 2003dg, SN 2003dr, and SN 2005E are grouped apart from other SN (Ib+Ic) because of their potentially distinct progenitor population;
(5) SN position coordinates in the host galaxy; (6) inside SDSS DR 8 imaging footprint;  (7) retrieved SDSS images collectively cover host galaxy without header issue; (8) host galaxy not detected (SN 2006jl (IIn); SN 2006lh (II); SN 2007fl (II); SN 2008bb (II); SN 2008it (IIn); SN 2009dv (IIP); SN 2009lz (IIP); SN 2009ny (Ib); PTF09gyp (IIb));
(9) no contamination from nearby bright stars; 
(10) no contamination from residual SN light, the sample used for photometry measurements; (10a) amateur discoveries in the targeted photometry sample; (11) an SDSS host fiber available with SPECTROTYPE=`GALAXY' and sufficient S/N to classify using BPT diagram; (12) no contamination from an active galactic nucleus (AGN) in SDSS spectrum;
(12a) fibers that are positioned on the host galaxy nucleus; (12b) where the difference between the fiber's host offset and the SN host offset is less than 3 kpc;
(12c) amateur discoveries in the spectroscopy sample;
Two SN-LGRB had $z < 0.08$ (SN 1998bw and SN 2006aj), and only the host of SN 2006aj was inside the SDSS DR8.
The middle and bottom sections of the Table correspond to the `Photometry' and the `Spectroscopic' samples, respectively, subsets of the SN remaining after the
`No Host Detected' criterion is applied.
}
\label{tab:selectionTargeted}
\end{deluxetable}  

\section{Data}
\label{sec:data}
The imaging component of the SDSS DR8 spans 14555 square degrees and consists of 53.9 s $u'g'r'i'z'$ exposures taken with the 2.5m telescope at Apache Point, New Mexico. Each 
frame consists of a 2048 $\times$ 1498 pixel array that samples a 13.5$'$ $\times$ 9.9$'$ field.  
The complementary fiber SDSS DR8 spectroscopic survey covers a 9274 square degree subset of the DR8 imaging footprint. Objects detected at greater than 5$\sigma$, selected as extended, and with $r'$-band magnitudes brighter than 17.77 comprise the main galaxy sample for spectroscopic targeting. When the $r'$-band 3$"$ fiber magnitude is fainter than 19 magnitudes, fiber targets must meet additional criteria, and physical constraints limit adjacent fibers to be no closer than 55$''$ \citep{stra02} in a single fiber mask. Because of their large angular sizes, nearby galaxies were often `shredded' into multiple objects
by the SDSS object detection algorithm [see Fig. 9 of \citet{bl05}], and many of these galaxies were targeted in multiple locations with fibers. 
Wavelength coverage of the SDSS spectrograph extends from 3800 to 9200 \AA.  Exposures typically are a total of 45 minutes taken in three separate 15 minute exposures.

We admit only the spectra that the SDSS pipeline classifies as a galaxy (SPECTROTYPE=`GALAXY'), a step that includes normal galaxies and Type 2 AGN but removes QSOs and Type 1 AGN.



\section{Sample}
\label{sec:sample}

We assemble our SN samples from discoveries collected in the Asiago catalog \citep{ba99} through 2010 November 6 and 72 Palomar Transient Factory (PTF) core-collapse SN discoveries from March 2009 through March 2010 \citep{arcavi10}. Eight of the SN in the Asiago catalog (all Type II SN) are also among the \citet{arcavi10} PTF SN
(IAU/PTF: 09ct/09cu; 09bk/09t; 09bj/09r; 09bl/09g; 09ir/09due; 09nu/09gtt; 10K/09icx; 10Z/10bau). 
Table~\ref{tab:selectionTargeted} shows the criteria we describe in this Section on the galaxy-impartial and targeted SN samples. 

\subsection{Excluding SN Contamination}

We consider images taken during the period from 3 months before to 12 months after discovery to be potentially contaminated by SN light.
\citet{arcavi10} do not report the exact discovery dates of PTF SN, so, for these SN, the contamination window begins three months before the
start and ends 12 months after the completion of the search period.
 
To assemble SDSS frames of each SN host, we first queried the SDSS SkyServer for any frames within 9.75' of the host galaxy center.
If none was available without possible contamination from SN light (even with partial coverage not including from the field center), we instead assembled and constructed mosaics from potentially contaminated images. 
Such mosaics were used only to measure the deprojected offsets of SDSS spectroscopic fibers and the SN site in each host galaxy. 
The header of each SDSS image provides keywords that
define a tangent plane projection (TAN) which maps 
coordinates on the sky to pixel coordinates. We discarded the small number 
of images retrieved from the SDSS server that lacked the keywords.


\begin{deluxetable*}{lcccccccc}
\tablecaption{SN Luminosity Functions}
\tablecolumns{8}
\tablehead{\colhead{Survey}&\colhead{Ia}&\colhead{II}&\colhead{IIn}&\colhead{IIb}&\colhead{Ib}&\colhead{Ic}&\colhead{Ic-BL}&\colhead{Ib+Ic}}
\startdata
LOSS&-18.49 (0.76)&-16.05 (0.17)&-16.86 (0.59)&-16.65 (0.40)&-17.01 (0.17)&-16.04 (0.31)&...&-16.09 (0.23)\\
P60&...&...&...&...&-17.0 (0.7)&-17.4 (0.4)&-18.3 (0.6)&...
\enddata
\tablecomments{Estimates of the mean luminosities of the SN types. The standard deviation of the luminosity function is shown in parentheses.
The LOSS (\citealt{li11a}) and the P60 (\citealt{drout11}) samples, respectively, are constructed differently, but differences between the mean luminosities of the 
SN species should be approximately consistent for these surveys. 
\citet{li11a} favor a much larger difference between SN Ib and SN Ic luminosities than that found by \citet{drout11}. SN Ic-BL may be more intrinsically luminous than SN Ic. 
Luminosities above are before correction for extinction,  for studying SN detection efficiency. }
\label{tab:luminosities}
\end{deluxetable*}

\subsection{Spectroscopic Classes}

Our previous work has shown that SN Ic are more strongly associated with bright regions in their host galaxies' $g'$-band light than are SN Ib (\citealt{kel08}), 
indicating that they have a distinct progenitor population, so we group them separately in this analysis.
SN IIb and SN IIn subtypes are excluded from the ``SN II" sample.
A single SN with an associated long duration gamma-ray burst (LGRB), SN 2006aj, meets the sample criteria, but we consider it separately from SN Ic-BL discovered through their 
optical emission (which have no associated LGRB), as did \citet{mod08}.
Today's gamma-ray searches are not sensitive to normal SN explosions. 
 
From a comprehensive set of spectra, we update the classification of SN 2005az. This SN was discovered approximately seventeen days before maximum and spectroscopically classified three days after discovery as a SN Ic by \citet{quimby05}. The Nearby Supernova Factory (SNF), from a spectrum taken five days after discovery, suggested it was a Type Ib (\citealt{aldering05}).  Spectral cross correlation using the Supernova Identification code (SNID; \citealt{blondin07}), applied to 24 spectra taken by the CfA Supernova Group from approximately ten days before to twenty-five days after maximum, shows that it was a Type Ic explosion. 
 
We update the spectroscopic types of ten SN found in the Asiago catalog with reclassifications
from CfA spectra (Modjaz et al. 2012, in preparation) using SNID. 
We also use new spectroscopic classifications for two SN from \citet{sandersuntargeted}, based on
revisions by the authors of the original IAU circulars. 
The SN in our sample that have new classifications from these papers have footnotes in Tables \ref{tab:paptable1}~and \ref{tab:paptable2}.  

We exclude SN 2006jc, a peculiar SN Ib with narrow helium emission lines and an underlying broad-lined SN Ic spectrum (e.g., \citealt{fol07}; \citealt{past07}), from our SN Ib statistical sample. The helium emission may reflect the collision of ejecta with a helium-rich circumstellar medium. 

We group calcium-rich SN 2000ds, SN 2003dg, SN 2003dr, and SN 2005E separately from other SN (Ib+Ic) because of their potentially distinct progenitor population (\citealt{perets10}).

We exclude SN IIn imposters (e.g., \citealt{vandyk00}; \citealt{maund06}), a group which includes SN 1997bs, SN 1999bw, SN 2000ch, and SN 2001ac. 

\subsection{Classification of SN as Type IIb}

While spectra taken over several epochs are necessary to observe the spectroscopic transition that defines SN IIb, such follow up is not always available. 
Fortunately, the spectra of Type IIb SN similar to SN 1993J are sufficiently distinctive that cross correlation with spectroscopic templates (e.g., SNID), has been able to identify substantial numbers of explosions as Type IIb from a single spectrum. Although classifications based on a single spectrum may overlook examples of SN IIb, the Type IIb explosions they do identify should be reliable. 

\subsection{Classification of SN as Type Ib/c SN}

The Asiago catalog entries sometimes have less information than the IAU Circulars and published papers.  For example, SN 1997dq and SN 1997ef were listed in the Asiago catalog (as of November 2010) as ``Type Ib/c" while \citet{math01b} and  \citet{mazzali04} identified them as SN Ic-BL. 
Motivated by these examples, we searched the circulars to see whether additional information was available. 
Despite making note of the presence or absence of helium more than ten days after the explosion, some authors report a Type Ib/c classification.
Authors may feel that a SN Ib/c classification was sufficiently precise while, in other cases, they may have wanted to emphasize conflicting spectroscopic characteristics. An example of the latter is SN 2003A which was classified as a Type Ib/c by \citet{fili03b} who noted that ``[w]eak He I absorption lines are visible, but the overall spectrum resembles that of type-Ic supernovae." 

Classifications by the Nearby Supernova Factory (\citealt{aldering02}; \citealt{wv04}) reported in circulars include an unusually high percentage of Type Ib/c.  
The high fraction of SN Ib/c reported by the Nearby Supernova Factory survey is hard for us to assess without being able to see the spectra or use impartial classification techniques. We have therefore excluded SN discovered by the Nearby Supernova Factory from our statistical sample. 




\begin{deluxetable*}{lccccccc}
\tablecaption{Mean Redshifts for Each SN Type}
\tablecolumns{6}
\tablehead{\colhead{Survey Type}&\colhead{II}&\colhead{IIn}&\colhead{IIb}&\colhead{Ib}&\colhead{Ic}&\colhead{Ic-BL}}
\startdata
Galaxy-Impartial&0.042&0.044&0.033&0.041&0.035&0.045\\
Targeted&0.013&0.014&0.012&0.014&0.011&0.013
\enddata
\tablecomments{Mean redshifts of each SN type in the galaxy-impartial and targeted samples.} 
\label{tab:meanRedshifts}
\end{deluxetable*}

\subsection{Galaxy-Impartial and Targeted SN Surveys}

We measure the host galaxy properties of SN discovered by both \textit{targeted} surveys, which aristocratically 
discover almost all their SN in luminous galaxies, as well as \textit{galaxy-impartial} surveys, 
which democratically scan swaths of sky without special attention to specific galaxies. 
Galaxy-impartial surveys generally employ larger 
telescopes (e.g., the SDSS 2.5m; the PTF 1.2m) 
than targeted surveys (e.g., the KAIT 0.76m), have fainter limiting magnitudes, and image much greater numbers of low-mass galaxies. 
The SN harvested by galaxy-impartial surveys are found in host galaxies that are not apparently bright or nearby (and are not in bright galaxy catalogs). 
For example, in our sample, 34\% (45/133) of galaxy-impartial SN but only 3.4\% (13/387) of targeted SN have host galaxy masses smaller than $10^{9} M_{\odot}$.



\subsection{Identifying Galaxy-Impartial Discoveries}

We used the Discoverer column from the IAU classification\footnote{http://www.cfa.harvard.edu/iau/lists/Supernovae.html} to determine the provenance of each SN. 
There are relatively few galaxy-impartial discovery teams, because discovering substantial numbers of SN by impartially scanning the sky requires significant dedicated ob- 
serving time and investment in data processing. Any SN whose discovery team we did not identify as part of a galaxy-impartial search effort, including amateur discoveries, was 
considered a targeted discovery. 


Surveys that we considered galaxy-impartial are as follows: Catalina Real-Time Sky Survey and Siding Spring Survey \citep{djor11}, La Sagra Sky Survey, PAN-STARRS \citep{kaiser10}, Palomar Transient Factory \citep{law09}, ROTSE \citep{yost06}, ESSENCE \citep{mik07}, Palomar-Quest \citep{djor08},  SDSS-II \citep{sako05}, Supernova Legacy Survey \citep{astier06}, Supernova Cosmology Project \citep{perlmutter99}, Near Earth Asteroid Tracking Program \citep{pravdo99}, High-z Supernova Search \citep{riess98}, Experience de Recherche dÕObjets Sombres (EROS) \citep{hardin00}, Great Observatories Origins Deep Survey (GOODS) \citep{dick03}, Deep Lens Survey \citep{wittman02}, and, except for discoveries in targets IC342, M33, M74, M81, NGC 6984, and NGC 7331, the Texas Supernova Search (\citealt{quimbytexas05}). 


\subsection{SN Detection}

The control time $t$ is the total time period during which a SN with a specific light curve, luminosity, and extinction by dust along the line-of-sight 
would have been detected by a survey's observations of a given galaxy (e.g., \citealt{cappellaro99}; \citealt{leaman11}).
Extinction along the line-of-sight to potential explosion sites in each monitored galaxy as well as to the actual sites of SN explosions, however, is challenging to estimate. Therefore, control times are generally estimated
for apparent luminosity functions and magnitudes uncorrected for extinction, 
instead of for dust-free luminosity functions and magnitudes with a galaxy-by-galaxy correction for obscuring dust. 
The expectation value of the number of discoveries of type $T$ SN during a survey in the $i$th monitored galaxy
is,
\begin{equation}
\langle N^T_i \rangle =  r_i^T \times t_i^T, 
\end{equation}
where $r_i^T$ is the rate of and $t_i^T$ is the control time for type $T$ SN in the galaxy.
The probability of detecting $N_{i}^T$
type $T$ SN in the $i$th galaxy imaged by a survey follows a Poisson distribution, 
\begin{equation}
P(N^T_i) = {\rm Pois}(r_i^T \times t_i^T).
\end{equation}

The $i$th galaxy observed by the survey has 
properties that include, for example, the stellar mass $M_i$. 
We are interested in making inferences about how the 
rates of the SN types may depend upon galaxy properties. 
The statistical approach we take in this paper is to look for differences among the 
distributions of host properties of the reported SN of each type $S^T(M_i)$ (see Section~\ref{sec:statmeth}),
\begin{equation}
S^T(M_i) \approx {\rm Pois}(r_i^T \times t_i^T).
\end{equation}






Here we cannot estimate control times because we lack information about the contributing surveys. 
We do not explicitly model the effect of different possible control times as well as other potential effects
in our statistical analysis. 

In the following sections, however, we find evidence that control times may be sufficiently similar
that any differences do not dominate the results we find.
We consider the published luminosity functions,
place appropriate redshift upper limits on samples, and examine
the redshift distributions of galaxy-impartial SN discoveries.
Only differences among control times for SN species that correlate with galaxy properties
will introduce bias into our statistical analysis. 

The number of galaxies monitored by galaxy-impartial surveys
grows with the volume within the limiting redshift (i.e., $\propto z^3$). 
The number of galaxies monitored by targeted surveys, by contrast,
likely does not increase as quickly with redshift, because the generally high-mass targets are selected
from galaxy catalogs that have, for example, Malmquist bias. 



\subsection{Luminosity Functions, Light Curves, and Detection}

Recent measurements have compared the mean luminosities of the core-collapse spectroscopic
species, but any differences are not yet well constrained. 
\citet{li11a} (LOSS) and \citet{drout11} (Palomar 60$"$) measured the mean peak absolute magnitudes (before correction for 
host galaxy extinction) of core-collapse species, and these values are shown in Table~\ref{tab:luminosities}. 
\citet{drout11} found that SN Ic-BL are intrinsically brighter explosions, on average, than 
normal Type Ic explosions, but the LOSS sample included too few examples to corroborate a difference.  
Although \citet{li11a} found some evidence that SN Ib and SN Ic have different 
mean intrinsic luminosities, \citet{drout11} did not find a similar indication. 

Luminosity functions are broad, so the control time will not 
necessarily vary strongly with the mean SN luminosity. 
\citet{li11a} found, for example, that the SN (Ib+Ic) luminosity function has a standard deviation of 1.24 mag and that the
SN II luminosity function has a standard deviation of 1.37 mag.

\subsection{Redshift Upper Limits for Targeted and Galaxy-Impartial Samples}

We exclude SN discovered at redshifts where only SN Ib, SN Ic, and SN II with brighter-than-average luminosities are detected.
Targeted surveys generally employ smaller telescopes and have shallower limiting magnitudes, 
because their galaxy targets are nearby. 
We select LOSS and SDSS as the representative targeted and galaxy-impartial surveys respectively, 
because they are responsible for the greatest numbers of discoveries in each category. 

From luminosity functions and search limiting magnitudes, we estimate these upper redshift limits where detection efficiency falls below 50\% as 
 $z=0.023$ and $z=0.08$ for LOSS and SDSS-II, respectively.
We use -16 as the mean absolute SN magnitude, because 
\citet{li11a} measured mean absolute magnitudes for SN (Ib+Ic), \mbox{-16.09 $\pm$ 0.23} mag, and SN II, \mbox{-16.05 $\pm$ 0.15} mag.
For LOSS, which contributes 42\% of our targeted sample, \citet{leaman11} report a median limiting magnitude of 18.8 $\pm$ 0.5, 
corresponding to a detection limit of $z=0.023$ for SN (Ib+Ic) and SN II. 
LOSS survey observations are taken without a filter, and the total response function peaks in the \textit{R} band \citep{li11a}.

\citet{dilday10} report a $\sim$21.5 mag $r'$-band detection limit for the SDSS-II survey, which accounts for 33\% of the galaxy-impartial sample. 
For our sample of galaxy-impartial discoveries, the redshift upper limit is $z=0.08$, corresponding to the SDSS-II detection limit. 
The PTF, accounting for 32\% of galaxy-impartial SN, has a limiting \textit{R}-band magnitude of $\sim$20.8 mag which corresponds to 
an upper detection limit of $z=0.056$. 



Varying the upper redshift limit for galaxy-impartial and targeted SN discoveries (e.g., from $z=0.023$ to $z=0.02$ or $z=0.06$) does not alter the type-dependent trends we find. Table~\ref{tab:meanRedshifts} presents the mean redshifts for each SN type.

\subsection{Amateur Discoveries}
From the information available in IAU circulars, we separated
discoveries into those made by amateur astronomers, who generally use comparatively small telescopes, and by professional astronomers. 
Table~\ref{tab:selectionTargeted} lists the numbers of SN in each 
sample that were amateur discoveries. We present significance values
for several sample comparisons with and without amateur discoveries in Section~\ref{sec:results}.

\subsection{Spectroscopic Fiber Locations On Host Galaxy}

We identified SDSS spectroscopic fibers that targeted the galaxy nucleus by visually inspecting images and fiber positions. 
Oxygen abundance varies primarily with offset from the galaxy center, and we also determined which fibers have host offsets within 
3 kpc of the SN host offset. The numbers of fibers in each category are listed in Table~\ref{tab:selectionTargeted}.

\begin{figure*}[t]
\centering
\subfigure{\includegraphics[angle=0,width=3.25in]{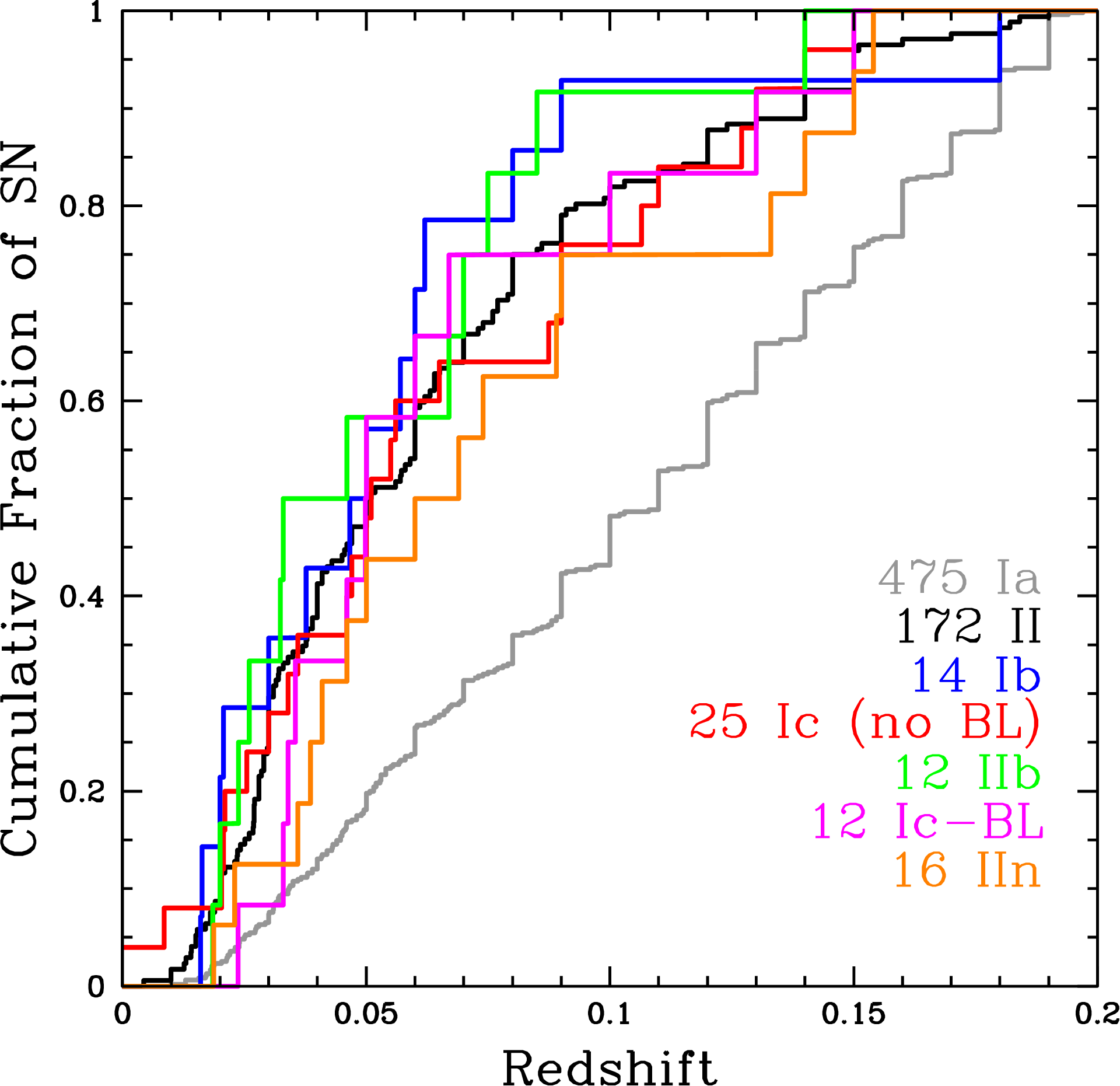}}
\subfigure{\includegraphics[angle=0,width=3.25in]{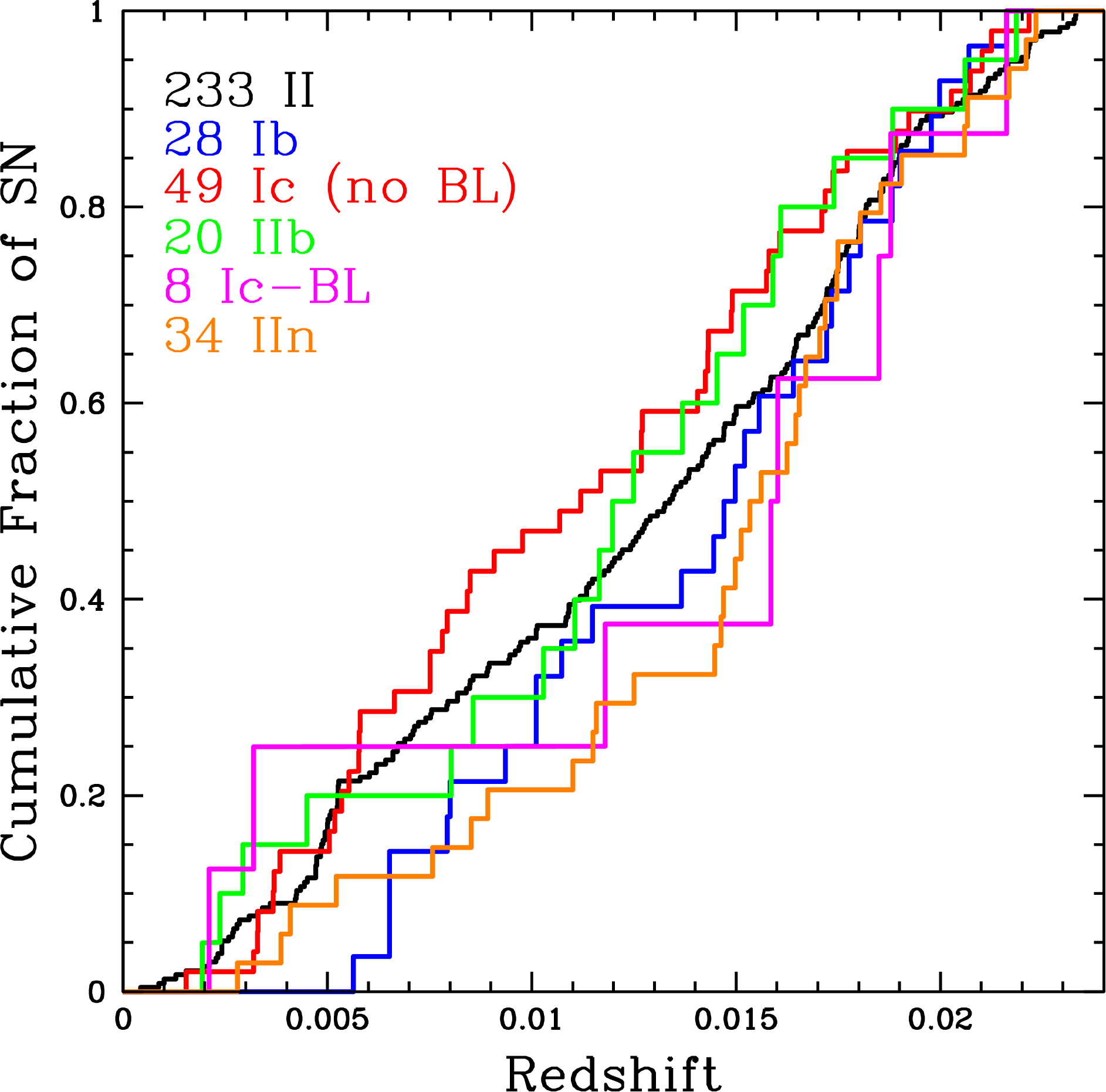}}
\caption{Redshifts of SN discovered by galaxy-impartial surveys to $z=0.2$ (left) and of SN discovered by targeted surveys in our sample (right). 
To increase the size of the sample, the left plot includes discoveries to $z = 0.2$, a higher upper limit than the $z = 0.08$ galaxy-impartial sample limit. 
Except for cosmic evolution, low-redshift galaxy-impartial surveys image the same galaxy populations at increasing redshift.
This plot provides no suggestion that the detection efficiency functions, $\eta(z,T,m_{\rm lim},C)$, for 
SN II, SN IIb, SN IIn, SN Ib, SN Ic, and SN Ic-BL vary differently with redshift, although the numbers of some species are small.
The plot on the right shows the redshift distributions of the SN in our sample discovered by targeted surveys.
Among the 15 possible two-sample comparisons, the most extreme difference is between the SN Ic and SN IIn distributions and has \mbox{$p=\icnoblXiinztar$}.
} 
\label{fig:recessional}
\end{figure*}

\section{Testing For Detection-Related Systematics}
\label{sec:systematics}

\subsection{Comparing Detection Control Times Indirectly with Galaxy-Impartial SN Discoveries}

Galaxy-impartial searches do not target specific galaxies, so 
their discovery rate may be expressed in terms of the rate of 
SN per unit volume, 
\begin{equation}
\langle N^T_{z_i} \rangle = r_{z_i}^T  \times V_{z_i} \times t_{z_i}^T,
\label{eqn:impartial}
\end{equation}
where $r_{z_i}^T$ is the type T SN rate per unit volume, $V_{z_i}$ is the volume within the survey field-of-view, and $t_{z_i}^T$ is the control time for type $T$ SN that explode in the
$z_i$ redshift bin (e.g., $0.01 < z < 0.015$).

The power-law form of the Schechter luminosity function (\citealt{sch76}) cuts off exponentially at the characteristic absolute magnitude, $M_*$.
The AGN and Galaxy Evolution Survey (AGES) found that $M_*$
becomes $\sim$0.2 mag brighter between $z=0$ and $z=0.2$ (\citealt{cool12}). 
\citet{moustakas11} reported, also from AGES spectroscopy, that 
the mean gas-phase metallicity of galaxies decreases by only 
$\sim$0.02-0.03 dex from $z=0.1$ to $z=0.2$. 
Given this evidence for relatively modest changes in the galaxy population, 
we may reasonably expect the fractional representation of each SN type among SN that explode to change only modestly to \mbox{$z$ $\approx$ 0.2},
\begin{equation}
\frac{r_{z_i}^A}{r_{z_i}^B} \approx  \frac{r_{z_j}^A}{r_{z_j}^B},
 \label{eqn:fraction}
\end{equation}
where $z_i$ and $z_j$ are redshifts less than 0.2 and $A$ and $B$ are distinct spectroscopic classes of SN. 
The relationships in Equations \ref{eqn:impartial} and \ref{eqn:fraction} suggest that 
changes in the ratio between the numbers of detected SN of two species with increasing
redshift depends primarily upon changes in the control times for the species,

\begin{equation}
\frac{d}{dz} \frac{\langle N^A \rangle}{\langle N^B \rangle} \approx \frac{r^{A}}{r^{B}} \frac{d}{dz} \Bigg( \frac{t^A}{t^B} \Bigg)
\end{equation}



The left panel of Figure~\ref{fig:recessional} plots the cumulative redshift distributions of core-collapse SN as 
well as Type Ia SN with $z<0.2$ reported to the IAU by galaxy-impartial surveys 
We extend the redshift upper limit beyond the $z=0.08$ galaxy-impartial limit to compile a larger sample of SN discoveries. 
This plot suggests that surveys may have similar control times for SN II, SN IIb, SN IIn, SN Ib, SN Ic, and SN Ic-BL with increasing redshifts, although the numbers of some species are small.
The Type Ia SN are, however, discovered at greater redshifts than the core-collapse species because of greater intrinsic brightness.
This test may only be sensitive to strong differences among control times, however, and we expect that a more complete analysis based on detailed knowledge of each survey and improved constraints on 
SN luminosity functions will find differences among the control times for the core-collapse spectroscopic types.

The control times for targeted surveys are likely to have similar behavior from $z=0$ to $z=0.023$ (where $\eta=0.5$) as those for galaxy-impartial surveys from $z=0$ to $z=0.08$ (where $\eta=0.5$).
Targeted surveys generally have shallower limiting magnitudes, but their galaxy targets are also at smaller distance.



\subsection{Redshift Distributions of Targeted SN Discoveries}

The redshift distributions of targeted SN, shown in the right panel of Figure~\ref{fig:recessional}, additionally depend on the set of galaxy search targets as well as type-dependent host galaxy preferences. 
A Malmquist effect in galaxy catalogs (e.g., the NGC) used to
select targets (\citealt{leaman11}) means that 
more distant monitored galaxies are, on average, more luminous, and likely more metal-rich.
The redshift distributions of the targeted SN samples in Figure~\ref{fig:recessional}
show greater differences than those for galaxy-impartial SN samples. 

\subsection{Detection Related Systematics}



Here we have only discussed possible systematic errors associated with 
differences between the luminosity functions of the core-collapse species and 
the redshift dependence of the galaxies monitored by SN searches. 
In Section~\ref{sec:tests}, we list additional potential sources of systematic
error including fiber targeting and spectroscopic classification. 


\section{Methods}
\label{sec:methods}

\subsection{SDSS Imaging Processing}

The WCS provided in SDSS DR8 {\bf frame*.fits} headers is a TAN approximation to the {\bf asTrans*.fits} full astrometric solution and has subpixel accuracy (private communication; M. Blanton), an improvement over the astrometric solution provided in DR7 {\bf fpC*.fit} headers.
We used SWarp \citep{bert02} to register and resample SDSS images of each host galaxy to a common pixel grid and coadded SDSS images in the $u'g'r'i'z'$
bands. 
The SDSS DR8 {\bf frame*.fits} images also feature an improved background subtraction (in comparison to DR7 {\bf fpC*.fit} images).  The DR8 background level is estimated from a spline fit across consecutive, adjacent frames from each drift scan, after masking objects in each image.

\subsection{Galaxy Photometry and Stellar Mass}

Host galaxy images were used to measure the color of the stellar population
near the site of each SN, estimate each host's stellar mass, a proxy for chemical enrichment (e.g., \citealt{tre04}), and 
compute the deprojected host offsets of SN and SDSS spectroscopic fibers.
 
The SExtractor measurement MAG\_AUTO, corresponding to the flux within 2.5 Kron (\citeyear{kron80}) radii, was used to estimate host galaxy stellar mass-to-light ratio ($M/L$) through fits with spectral energy distributions (SEDs) from PEGASE2 (\citealt{fi97}, \citeyear{fi99}) stellar population synthesis models using the appropriate SDSS instrumental response function.
An estimate of the stellar mass was then computed: $M = M/L \times L$ where $L$ is the galaxy's absolute luminosity. 
See \citet{kel10} for a detailed description of the star formation histories used. 

\subsection{Color Near Explosion Site}

We then estimated the host galaxy's color near the SN location using two techniques. 
The first and more simple method was to extract the $u'g'r'i'z'$ flux inside of a circular
aperture with 300 pc radius centered on the SN location, after subtracting the
median of the peripheral background regions. 
While this technique is straightforward, a small number of apertures centered at the sites of SN at large host offsets or found in low-luminosity hosts had low S/N, especially in the $u'$ and $z'$ bands.
The primary intent of the second method was to obtain higher S/N 
$u'$-band flux measurements near the sites of SN, in particular near the sites of the 
SN with faint hosts. To identify $g'$-band pixels with S/N $>$ 1 associated with each host galaxy, we used SExtractor to generate a
segmentation map of each image, which identifies the pixels associated with each object. 
We adjusted the SExtractor settings so that the segmentation map included
only pixels with S/N $>$ 1 (i.e., DETECT\_THRESH=1).
The 20 pixels closest to the SN location contained in the $g'$-band segmentation maps
define the aperture for measurements of $u'$-$z'$ color and $u'$-band surface brightness. 

The aperture generally consists of the 20 pixels on the CCD array closest to 
the SN position (i.e., within a circle of radius $\sim$1$''$). Only for twenty of the \bhnumber~SN is the average distance between aperture pixels and the
explosion site greater than 0.8$"$. 
Excluding measurements where the average pixel is more than $0.8"$ from the explosion coordinates
does not affect the distributions we plot in this paper. The median uncertainties of the measured $u'g'r'i'z'$ magnitudes are 
0.11, 0.06, 0.06, 0.06, and 0.13 magnitudes, respectively. 
We correct for Galactic reddening using the \citet{sch98} dust maps.

The reported SN positions come from a variety of sources that may rely on 
catalogs (e.g., USNO B1, 2MASS, or GSC) with more modest astrometric accuracy
than available for SDSS images. Repeated photometric follow-up, available for some SN, can also be useful for improving the  
accuracy of explosion coordinates (e.g., \citealt{hicken12}).


We use KCORRECT \citep{bl07} to estimate rest-frame $u'g'r'i'z'$ magnitudes from the fluxes we measure. 
KCORRECT fits the input measured fluxes with a model for the rest-frame SED, and it uses this
SED to estimate the rest-frame $u'g'r'i'z'$ magnitudes for each galaxy.
The estimated rest-frame $u'-z'$ color, for example,
therefore depends upon the full set of measured observer-frame 
$u'g'r'i'z'$ magnitudes that inform the SED model.
Consequently, even if $u'$ and $z'$ measurements are noisy, the estimated rest-frame 
 $u'-z'$ color may be informative, given constraints from less noisy measured $g'r'i'$ fluxes.




\subsection{Selecting SDSS Spectroscopic Fibers}

To identify SDSS fibers that coincide with a host galaxy, we searched inside a catalog available online from an MPA-JHU collaboration\footnote{http://www.mpa-garching.mpg.de/SDSS/DR8/} for fibers that fell within an aperture with radius $(1.65/z)"$ placed on the host center and with redshifts that agree with that of the SN.  For an object in the Hubble flow, this angle corresponds to a physical distance of approximately 34 kpc. At $z=0.03$, for example, this radius subtends a 55$"$ angle. If the $g'$-band SExtractor segmentation map ID at the fiber location was the same as the ID of the SN host galaxy, the fiber was considered a match to the galaxy after a visual check. 
The deprojected normalized offset of the fiber was then calculated by computing the offset at each pixel in the 3$"$ fiber aperture and averaging these offsets weighted by each pixel's $g'$-band counts.

\subsection{AGN Activity}
\label{sec:agn}

Only the spectra classified by the SDSS pipeline as a galaxy spectrum (SPECTROTYPE=`GALAXY') enter our analysis, a 
restriction that excludes quasi-stellar objects (QSO) and Type 1 active galactic nuclei (AGN) whose continua have significant non-stellar contributions. 
The SDSS `galaxy' class, however, includes spectra with emission line strength ratios characteristic of 
Type 2 AGN and low ionization nuclear emission regions (LINERs).
The emission line patterns associated with AGN activity are significantly degenerate with variation in oxygen abundance, so
AGN line ratios preclude metallicity measurements.

We use the classifications of fiber spectra as star forming, low S/N star forming, composite, AGN, or low S/N AGN made available by the MPA-JHU group following \citet{brinchmann04}. 
That analysis uses each spectrum's position on the \citet{baldwin81} (hereafter BPT) diagram of [\ionpat{O}{iii}] $\lambda$5007/H$\beta$ and [\ionpat{N}{ii}] $\lambda$6584/H$\alpha$ line ratios.

\subsection{Extinction Estimated from Balmer Ratios}

From the fiber spectra (closest in deprojected offset to the SN sites), we estimate host galaxy reddening $A_{V}$ using the Balmer decrement (H$\alpha$/H$\beta$), assuming the $R_{V}$=3.1 \citet{cardelli89} extinction law. 
Following \citet{ost89}, we assume a Case B recombination ratio of 2.85 when spectra are classified as star forming or low S/N star forming and a ratio of 3.1 when spectra are classified as composite, AGN, or low S/N AGN.



\begin{figure*}[t]
\centering
\subfigure{\includegraphics[angle=0,width=6.5in]{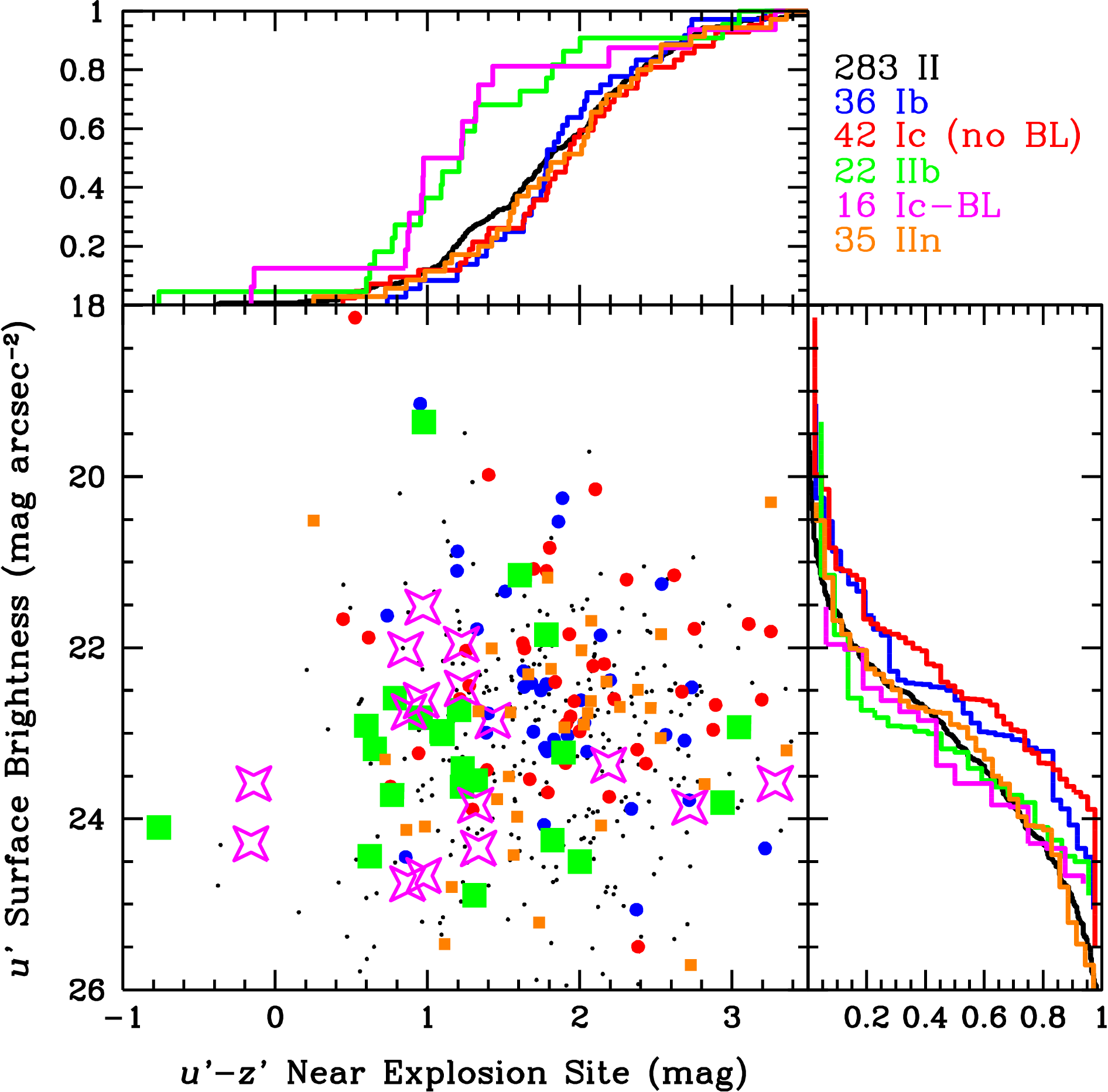}}

\caption{Host galaxy $u'-z'$ color versus $u'$ surface brightness near SN location. The top panel plots the fraction of
SN of each type with environments bluer than the horizontal axis coordinate, and the right panel plots the fraction of SN of each type whose 
environments have higher $u'$-band surface brightness than the vertical axis coordinate. 
SN IIb environments are bluer than SN Ib, SN Ic, and SN II environments (\mbox{$p=\iibXibuzkc$}, \iibXicnobluzkc, and \iibXiiuzkc, respectively), and SN Ic-BL environments are bluer than those of SN Ib, SN Ic, and SN II (\mbox{$p=\icblXibuzkc$}, \icblXicnobluzkc, and \icblXiiuzkc, respectively). 
SN Ib and SN Ic explode in regions with higher $u'$-band surface brightness than do SN II (\mbox{$p=\ibXiiukc$}~and \icnoblXiiukc, respectively), and SN Ic sites have higher $u'$-band 
surface brightnesses than SN Ic-BL locations (\icnoblXicblukc).
The aperture is the 20 host pixels with S/N $>$ 1 in $g'$ band nearest the SN location. 
}
\label{fig:uzcolor}
\end{figure*}

\begin{figure*}[t]
\centering
\subfigure{\includegraphics[angle=0,width=6.5in]{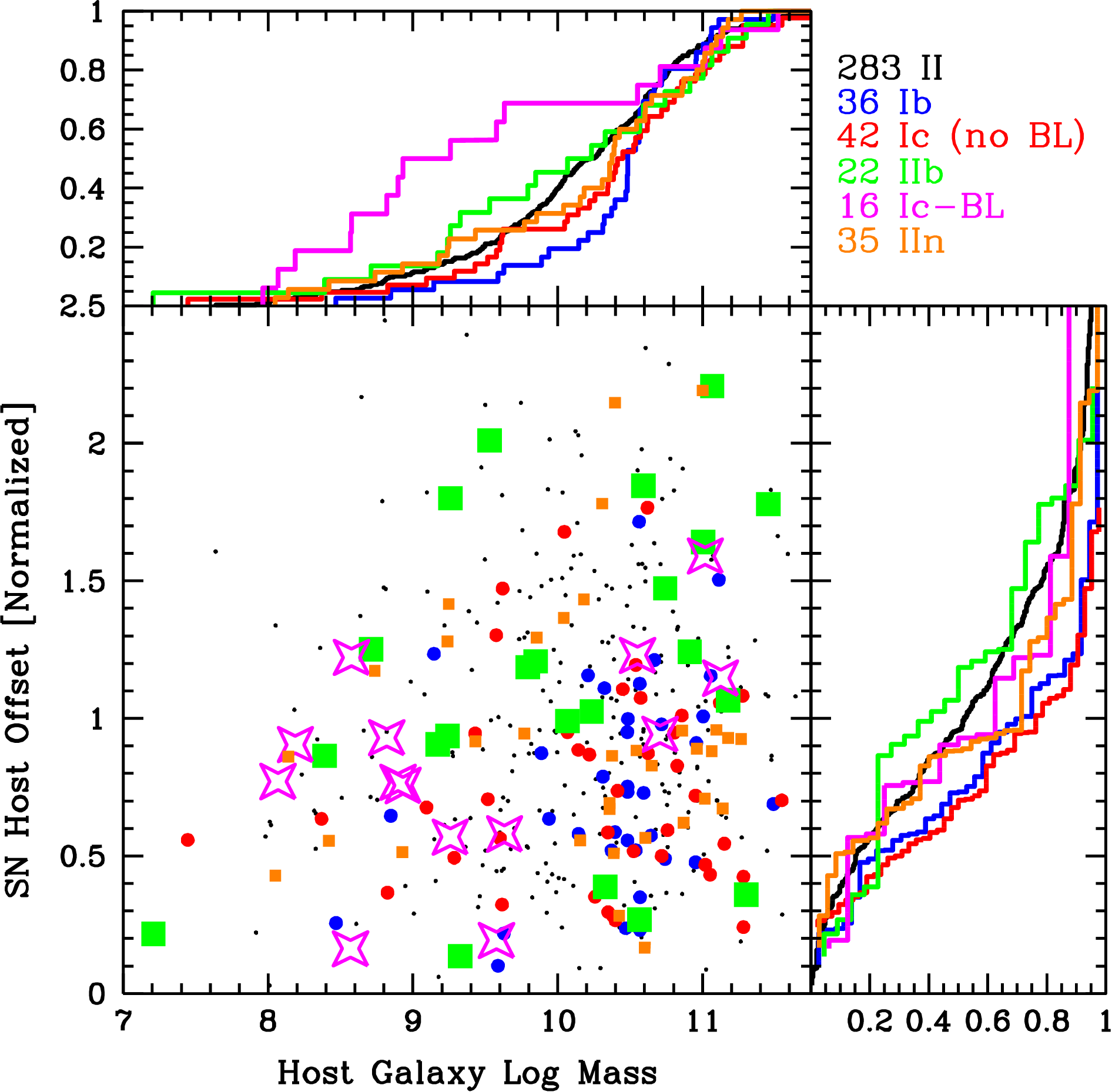}}
\caption{Stellar masses of host galaxies versus SN host offset, deprojected and normalized by host $g'$-band half-light radius. The top panel plots the fraction of each SN type with hosts less massive than the horizontal axis coordinate, 
and the right panel plots the fraction of SN of each type whose 
host offsets are greater than the vertical axis coordinate.  
SN Ic-BL are found in significantly less massive galaxies than are the SN Ib, SN Ic, or SN II (\mbox{$p=\icblXibLogM$}, \icblXicnoblLogM, and \icblXiiLogM, respectively).  Host galaxy stellar masses are estimated from PEGASE2 fits to $u'g'r'i'z'$ host magnitudes. A two-sample KS test finds evidence (\mbox{$p=\ibXiibsnrallhighmass$})~that SN IIb explode at larger host offsets than SN Ib, among the SN discovered in galaxies with log M $> 9.5$. SN Ic explode closer to their host centers than SN II  (\mbox{$p=\icnoblXiisnrall$}).
}
\label{fig:masses}
\end{figure*}



\subsection{Metallicity and Specific SFR Measurements}

Our analysis uses both (a) abundances and specific star formation rate estimates available from the MPA-JHU collaboration for SDSS fiber spectra and (b) abundances we compute using the \citet{pettini04} metallicity calibration. 
We only use galaxies with S/N$>$3 H$\beta$, H$\alpha$, [\ionpat{N}{ii}] $\lambda$6584, and [\ionpat{O}{iii}] $\lambda$5007, as designated by the MPA-JHU analysis.
For abundance measurements, we only analyze spectra classified as star forming. For specific SFR estimates, we use star forming, composite, and AGN spectra.


\subsubsection{MPA-JHU Metallicity and Specific SFR}
\label{sec:mpajhu}

To extract an oxygen abundance and specific SFR from a spectrum, the MPA-JHU collaboration first uses \citet{char01} stellar population synthesis and photoionization models to calculate an extensive library of line strengths spanning potential effective gas parameters including gas density, temperature, and ionization as well as the dust-to-metal ratio.
Then galaxy [\ionpat{O}{ii}], H$\beta$, [\ionpat{O}{iii}], H$\alpha$, [\ionpat{N}{ii}], and [\ionpat{S}{ii}] optical nebular emission lines are fit simultaneously with the library and used to compute metallicity and specific SFR likelihood distributions.  Here we use the median of these distributions as the oxygen abundance and specific SFR estimates. 
We refer to the metallicity estimates as T04 oxygen abundances, in reference to Tremonti et al. 2004 who 
employed the MPA-JHU values. 

When emission lines show AGN patterns, metallicity estimates are not possible from emission lines.
For these spectra, the MPA-JHU group uses the strength of the 4000 \AA~break [see Figure 11 of \citet{brinchmann04}] and the ratio
H$\alpha$/H$\beta$ to estimate the specific SFR\footnote{http://www.mpa-garching.mpg.de/SDSS/DR7/sfrs.html}. 
To calibrate the 4000 \AA~break as a specific SFR proxy,
star-forming spectra are placed into bins according to the strength of the 4000 \AA~break as well as the H$\alpha$/H$\beta$ ratio, a proxy for interstellar extinction. The galaxies in each bin are then used to compute the expected specific SFR for each set of parameters. 

\citet{kauffmann03agn} found that a sample of SDSS spectra of Type 2 AGN with median $z \approx 0.1$ and
selected according to the criteria we apply here show no evidence of a significant superposed 
AGN continuum. 
\citet{schmitt99} showed that AGN emission rarely accounts for more than 5\% of the continuum
of nuclear spectra of nearby galaxies with Type 2-patterned emission lines. 

\subsubsection{Pettini and Pagel Metallicity}

Since we have no prejudice about which emission-line method is most correct, we have also computed abundances using the \citet{pettini04} (hereafter PP04) prescription.  
This is based on the relative line
strengths of H$\beta$, H$\alpha$, [\ionpat{N}{ii}] $\lambda$6584, and [\ionpat{O}{iii}] $\lambda$5007, after correcting for dust emission. The PP04 indicator relies on lines relatively close in wavelength, reducing its sensitivity to uncertainty in the extinction correction and does not require the 
[\ionpat{O}{ii}] $\lambda3727$ line, which falls beyond the blue sensitivity of the SDSS spectrograph for objects at $z < 0.02$.

Our measurements trace the \citet{kew08} PP04 mass-metallicity relation of SDSS galaxies
when stellar mass is plotted against nuclear metallicity for galaxies in the Hubble flow ($z > 0.005$). 


\subsection{Comparison of Host Abundance Proxies}
\label{sec:proxies}
Oxygen abundances measured from fibers centered on the host galaxy nucleus
are, on average, only 0.01 dex (T04) greater than the abundance inferred from the
stellar mass with the \citet{tre04} \textit{M-Z} relation, with a scatter of 0.14 dex.
If we instead select fibers closest in host offset to SN explosion sites, 
spectroscopic abundances are 0.053 dex (T04) less than abundances estimated from stellar masses with a scatter of 0.16 dex.

\begin{figure}[t]
\centering

\subfigure{\includegraphics[angle=0,width=3.25in]{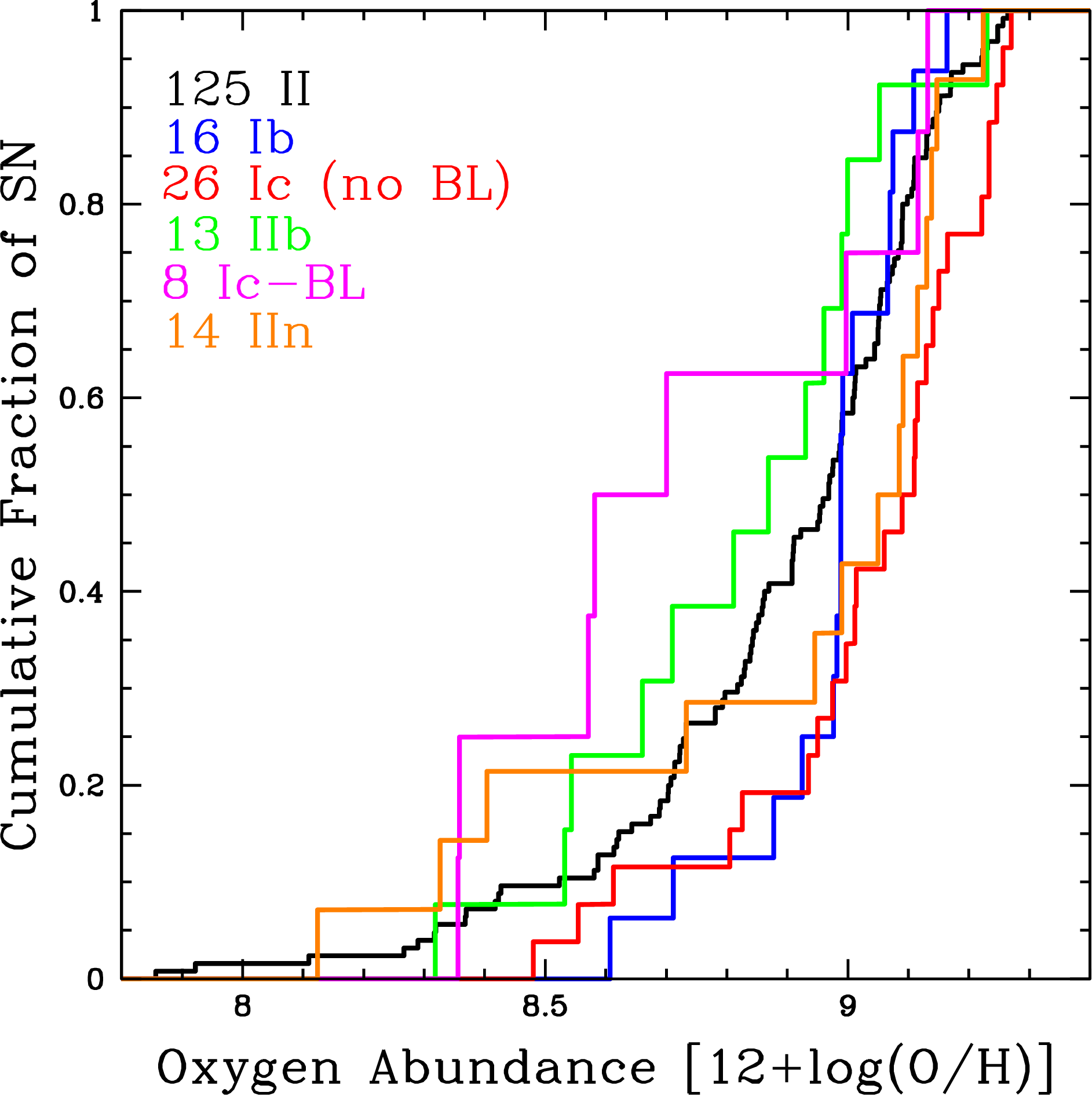}}

\caption{Host oxygen abundance measured from SDSS 3$"$ fiber spectrum with host radial offset most similar to that of SN explosion site. 
While \citet{tre04} spectroscopic abundances are plotted, we also measure abundances using the \citet{pettini04} calibration.
Even when we consider only SN discovered by galaxy-impartial surveys, 
we find a statistically significant difference between the SN Ic-BL and the SN Ic host abundance distributions
(\mbox{$p=\icblXicnoblclosetohfouruntar/\icblXicnoblcloseppohfouruntar$}, respectively for the T04/PP04 calibrations).
When we consider only SN discovered by targeted surveys, 
we find a statistically significant difference between the SN IIb and the SN Ib host abundance distributions
for one of two abundance diagnostics
(\mbox{$p=\iibXibclosetohfourtar/\iibXibcloseppohfourtar$}, respectively for the T04/PP04 calibrations).
Evidence for a difference between the SN IIb and SN Ib 
host distributions strengthens when all SN discoveries are considered (\iibXibclosetohfour/\iibXibcloseppohfour). 
}
\label{fig:abundances}
\end{figure}



\begin{figure}[t]
\centering
\subfigure{\includegraphics[angle=0,width=3.25in]{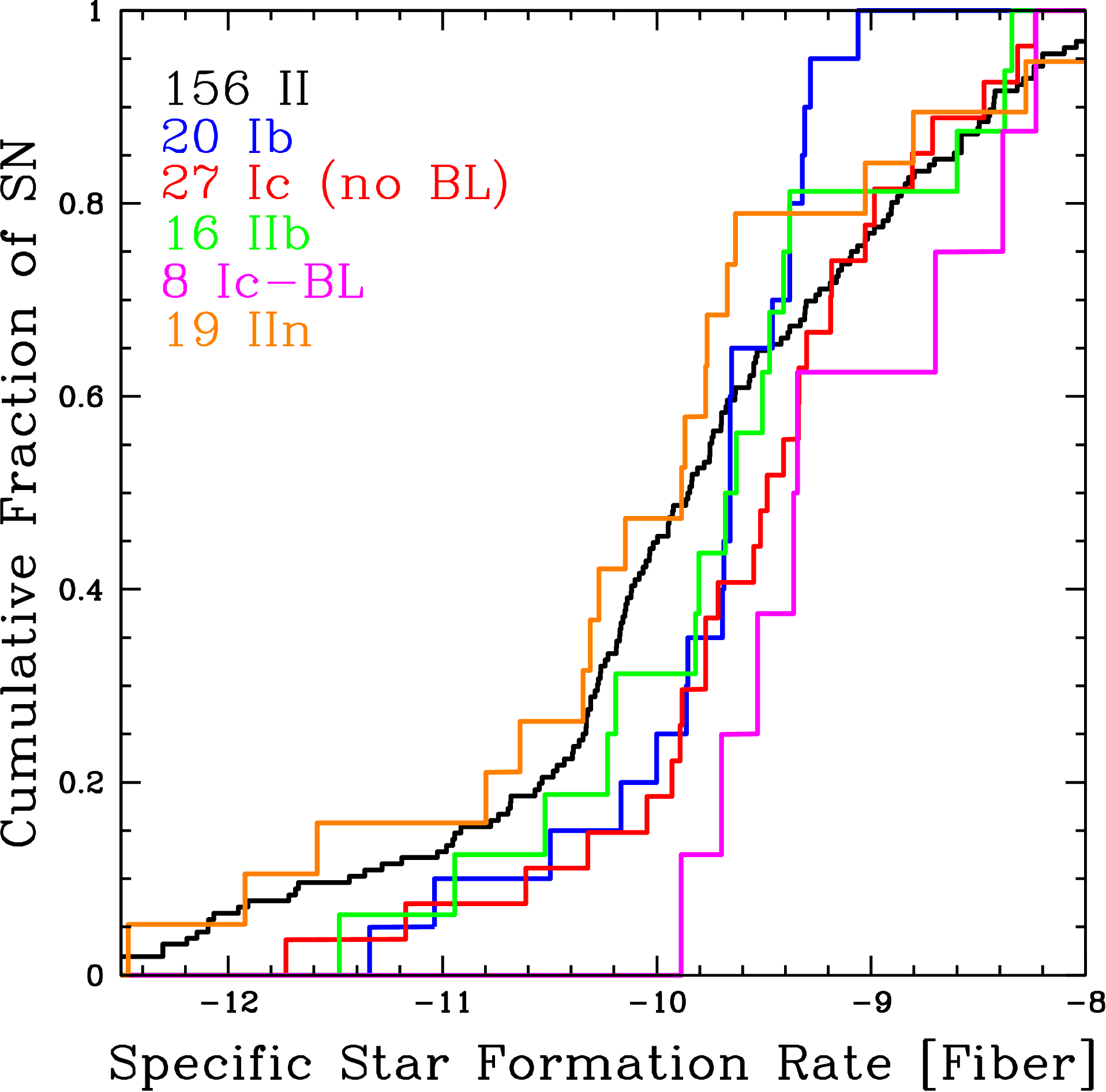}}

\caption{Host specific SFR estimated from SDSS 3$"$ fiber spectrum with host radial offset most similar to that of SN explosion site. The sequence of  the spectroscopic classes, arranged in order of the loss of the progenitor's outer hydrogen and helium envelopes (i.e., SN II, SN IIb, SN Ib, SN Ic), exhibit increasing average host galaxy specific SFR (SFR M$_{\odot}^{-1}$ yr$^{-1}$), measured from SDSS fiber spectra. 
SN (Ib+Ic) hosts have greater specific SFR than SN II hosts (\mbox{$p=\ibicXiiSSFRhighsn$}).
SN Ic-BL hosts have greater specific SFR than SN II hosts (\icblXiiSSFRhighsn). SDSS fibers largely sample light within the the host galaxy half-light radius and are often centered on the host galaxy nucleus.  
}
\label{fig:ssfr}
\end{figure}

\begin{figure}[t]
\centering
\subfigure{\includegraphics[angle=0,width=3.25in]{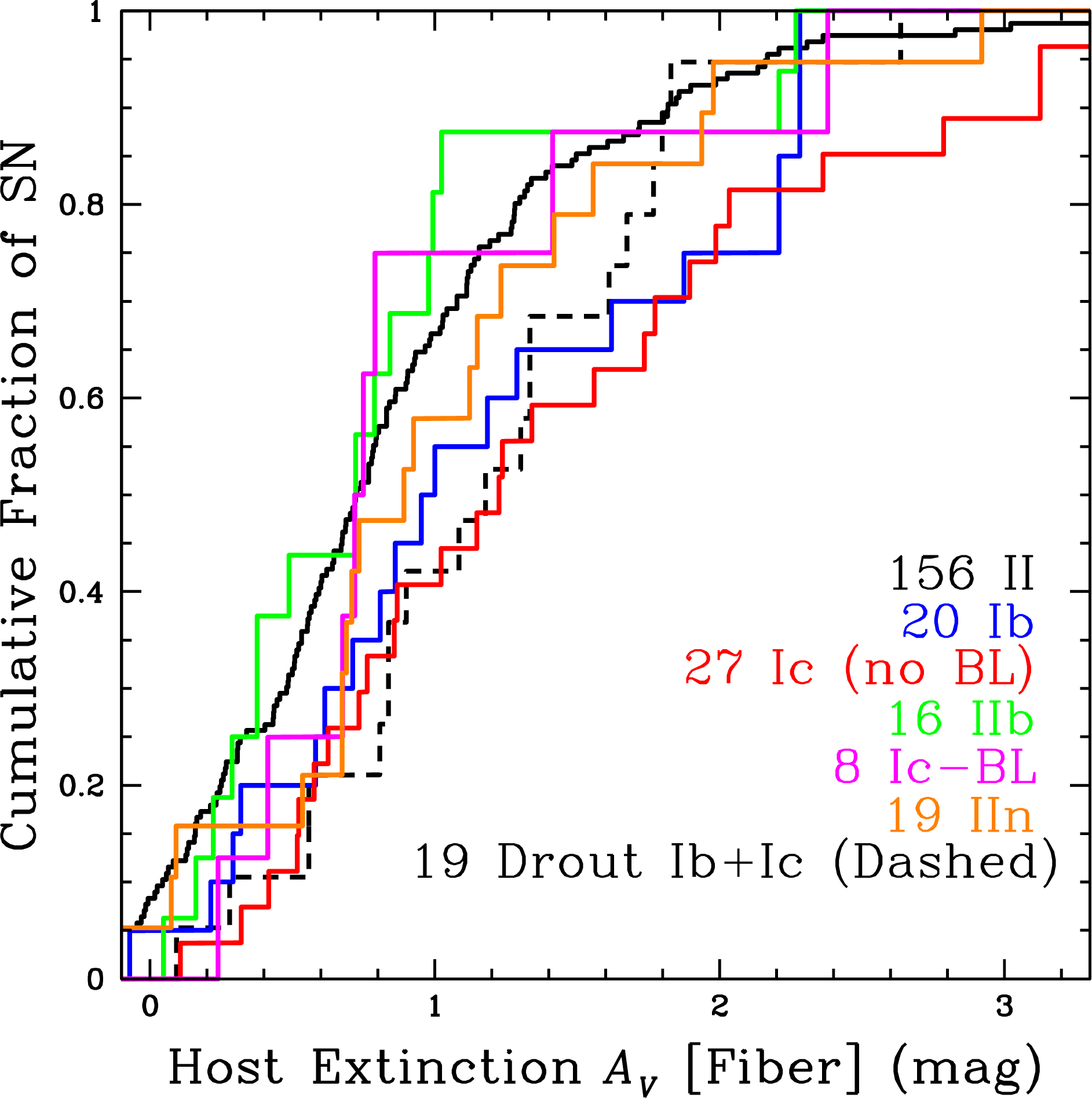}}
\caption{Host extinction estimated from 3$"$ SDSS fiber spectrum with host radial offset most similar to that of SN explosion site. 
There is significant evidence that SN IIb and SN II hosts have less internal extinction than SN (Ib+Ic) host galaxies (\mbox{$p=\iibXibicnoblAv$}~and \iiXibicnoblAv, respectively).
The \citet{drout11} SN (Ib+Ic) $A_V$ extinction values estimated along the line of sight to the SN from light curve color and shape are consistent
with the values we measure, although the spectroscopic fibers are largely not positioned at the SN site.  
}
\label{fig:abextinct}
\end{figure}

\begin{figure*}[t]
\centering
\includegraphics[angle=0,width=6.25in]{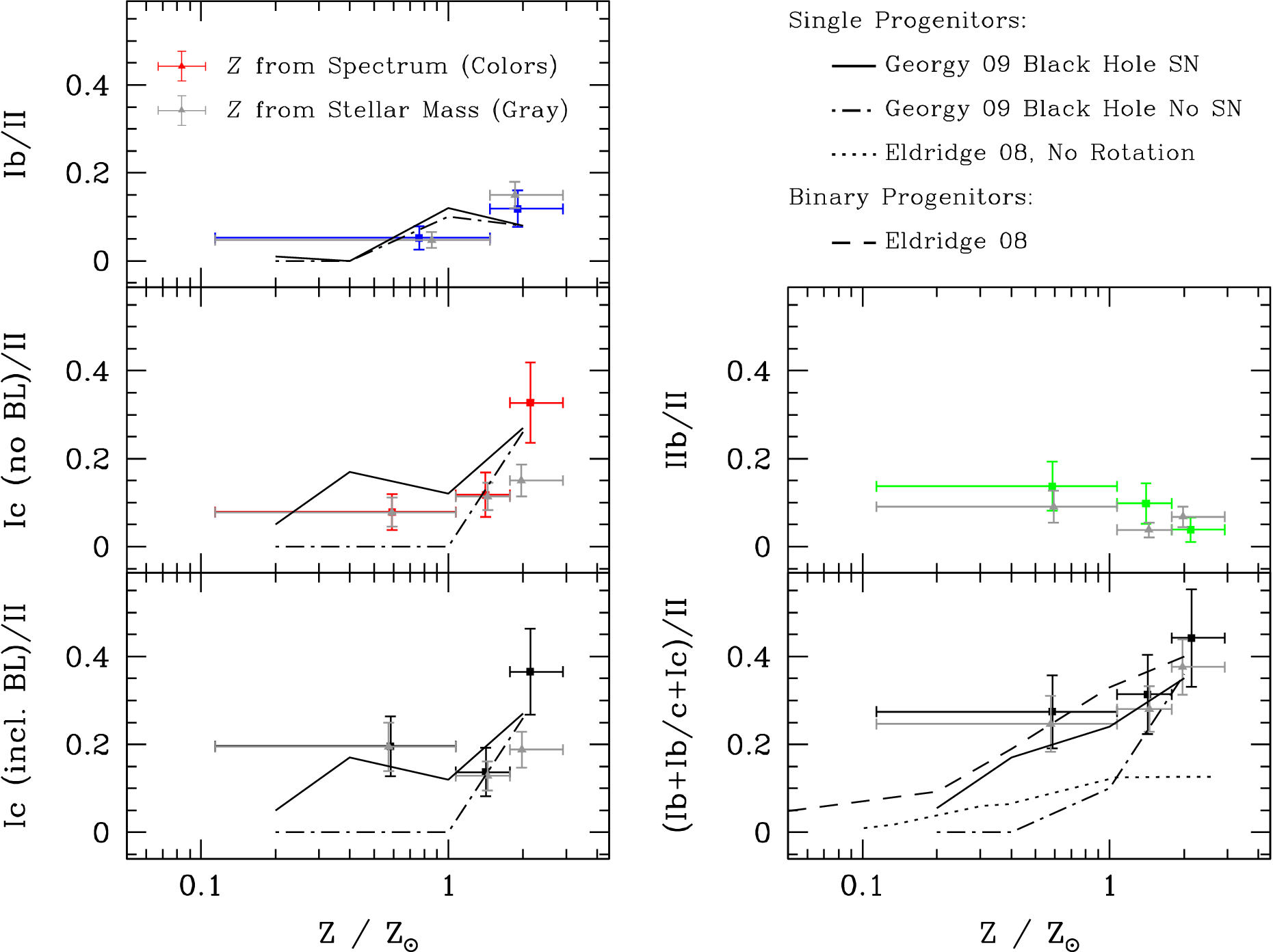}
\caption{
Ratio of stripped-envelope SN to SN II versus oxygen abundance (T04 calibration). 
The comparatively high fraction of SN (Ib+Ib/c+Ic) to SN II at subsolar metallicity in the right lower panel favors contributions from
a binary progenitor population or explosions even after collapse to a black hole.
Color points correspond to spectroscopic metallicity measurements, and gray points correspond to metallicities estimated from stellar masses using the \citet{tre04} mass-metallicity relation. 
The comparatively high fraction of SN II host fiber spectra with contamination from AGN activity (present only in massive, metal-rich galaxies) excludes a considerable fraction of metal-rich SN II 
host galaxies, inflating the apparent fraction of stripped-envelope SN in metal-rich galaxies (color points). Indeed, the stripped-envelope fraction is smaller using metallicities estimated from 
host galaxy stellar masses (gray) which do not suffer from an AGN selection effect.
Dashed line is \citet{eld08} prediction for binary progenitors; dotted line is \citet{eld08} prediction for non-rotating single progenitors; and solid and dash-dot lines are \citet{georgy09} predictions for single, rotating progenitors (where a minimum helium envelope of 0.6 M$_{\odot}$ separates SN Ib from SN Ic progenitors). Whether core collapse to a black hole can yield a SN explosion is not clear (e.g., \citealt{fry99}), especially if high angular momentum does not support an accretion disk (\citet{wo93}). The \citet{georgy09} solid line prediction is where core collapse to a black hole produces SN while the dashed-line prediction is where core collapse to a black hole yields no SN. Vertical error bars reflect Poisson statistics while horizontal bars reflect the range of metallicities in each bin with the position of the vertical bar corresponding to the mean Z in the bin. Here Z$_{\odot} = 8.86$ from \citet{delahaye10}. }
\label{fig:ratios}
\end{figure*}

\section{Statistical Method}
\label{sec:statmeth}
\subsection{Kolmogorov-Smirnov Statistic}

In the following sections, we test the null hypothesis that two samples are drawn from a single underlying distribution using the Kolmogorov-Smirnov (KS) test. The KS test statistic is defined as $D=\sup_x |F_1(x)-F_2(x)|$, the maximum difference between the samples' cumulative distribution functions, where $F_n(x)={1 \over n}\sum_{i=1}^n I_{X_i\leq x}$.  The KS distribution is the distribution of the test statistic $D$, given the null hypothesis that two distributions are identical. The $p$-value is the probability of observing a value of the test statistic, $D$, more extreme than the observed value of $D$ given the null hypothesis that the two samples are drawn from a single underlying distribution. Low $p$-values ($<5\%$) are significant evidence that the underlying distributions are distinct.

When two independent samples are drawn from the same distribution, there is, by definition, a 5\% random chance of obtaining a $p$-value less than 5\%.  If we were to make, for example, twenty comparisons among samples drawn from identical distributions, one misleading $p<5$\% difference would occur by chance on average. The number of independent comparisons we make in this paper should therefore be taken into account when comparisons yield $p$-values of modest significance ($p \approx5\%$). We note that the host properties we measure are correlated (e.g., host color and metallicity), so independent comparisons are fewer than the total number of comparisons. 

\section{Results}
\label{sec:results}

\begin{deluxetable*}{cccc}[b]  
\tablecaption{KS $p$-values for Combined Samples}
\tablehead{\colhead{Measurement}&\colhead{Figure}&\colhead{Samples}&\colhead{P-value}}
\startdata
$u'$-$z'$& \ref{fig:uzcolor} & Ic-BL vs. Ib, Ic, II&\icblXibuzkc, \icblXicnobluzkc, \icblXiiuzkc\\
... & ... & IIb vs. Ib, Ic, II&\ibXiibuzkc, \icnoblXiibuzkc, \iiXiibuzkc\\
$u'$ SB & \ref{fig:uzcolor} & Ib vs. Ic, II & \ibXiiukc, \ibXicnoblukc\\
... & ... &Ic vs. II & \iiXicnoblukc\\
log M & \ref{fig:masses} & Ic-BL vs. II, IIb, Ib, Ic & \icblXiiLogM, \icblXiibLogM, \icblXibLogM, \icblXicnoblLogM \\
... & ... &IIb vs. II, Ib, Ic & \iibXiiLogM, \iibXibLogM, \iibXicnoblLogM \\
... & ... & II vs. Ib, Ic, (Ib+Ic) & \iiXibLogM, \iiXicnoblLogM, \iiXibicnoblLogM \\
Offset & \ref{fig:masses} & Ic-BL vs. II, Ib, Ic & \icblXiisnrallhighmass, \icblXiibsnrallhighmass, \icblXibsnrallhighmass, \icblXicnoblsnrallhighmass \\
... & ... & IIb vs. II, Ib, Ic & \iibXiisnrallhighmass, \iibXibsnrallhighmass, \iibXicnoblsnrallhighmass \\
... & ... & II vs. Ib, Ic & \iiXibsnrallhighmass, \iiXicnoblsnrallhighmass \\
T04/PP04 & \ref{fig:abundances} & II vs. Ib, Ic & \ibXiiclosetohfour/\icnoblXiicloseppohfour, \icnoblXiiclosetohfour/\icnoblXiicloseppohfour \\
$<$ 3 kpc & ... & Ic-BL vs. Ib, Ic & \icblXibclosethreekpctohfour/\icblXibclosethreekpcppohfour, \icblXicnoblclosethreekpctohfour/\icblXicnoblclosethreekpcppohfour  \\
... & ... & II vs. Ib, Ic & \iiXibclosethreekpctohfour/\iiXibclosethreekpcppohfour, \iiXicnoblclosethreekpctohfour/\iiXicnoblclosethreekpcppohfour  \\
SSFR & \ref{fig:ssfr} & II vs. Ib, Ic, Ic-BL & \ibXiinobSSFRhighsn, \icnoblXiiSSFRhighsn, \icblXiiSSFRhighsn \\
... & ... & Ib vs. Ic & \ibXicnoblSSFRhighsn \\
$A_{V}$ & \ref{fig:abextinct} & (Ib+Ic) vs. IIb, II & \iibXibicnoblAv, \iiXibicnoblAv 
\enddata
\tablecomments{P-values from Kolmogorov-Smirnov two-sample comparisons that  include both targeted SN and galaxy-impartial SN discoveries.
The two rows below ``T04/PP04" show oxygen abundance statistics computed from spectra whose host offsets are within 3 kpc of
the SN host offset. 
The statistics comparing offsets includes only SN found in massive galaxies (log $M > 9.5$).
}
\label{tab:kscombo}
\end{deluxetable*}

\begin{deluxetable}{llcc}[b]  
\tablecaption{KS $p$-values for Different Samples}
\tablehead{\colhead{}&\colhead{Sample}&\colhead{IIb vs. Ib}&\colhead{Ic-BL vs. Ic}}
\startdata
\textbf{$u'$-$z'$}&\textbf{All SN}&\textbf{\iibXibuzkc~(\numiibuzkc, \numibuzkc)}&\textbf{\icnoblXicbluzkc~(\numicbluzkc, \numicnobluzkc)}\\
&Targeted&\iibXibuzkctar~(\numiibuzkctar, \numibuzkctar)&\icnoblXicbluzkctar~(\numicbluzkctar, \numicnobluzkctar)\\
&Impartial&\iibXibuzkcuntar~(\numiibuzkcuntar, \numibuzkcuntar)&\icnoblXicbluzkcuntar~(\numicbluzkcuntar, \numicnobluzkcuntar)\\
&No Amateur&\iibXibuzkcpro~(\numiibuzkcpro, \numibuzkcpro)&\icnoblXicbluzkcpro~(\numicbluzkcpro, \numicnobluzkcpro)\\
\textbf{log M}&\textbf{All SN}&\textbf{\iibXibLogM~(\numiibLogM, \numibLogM)}&\textbf{\icnoblXicblLogM~(\numicblLogM, \numicnoblLogM)}\\
&Targeted&\iibXibLogMtar~(\numiibLogMtar, \numibLogMtar)&\icnoblXicblLogMtar~(\numicblLogMtar, \numicnoblLogMtar)\\
&Impartial&\iibXibLogMuntar~(\numiibLogMuntar, \numibLogMuntar)&\icnoblXicblLogMuntar~(\numicblLogMuntar, \numicnoblLogMuntar)\\
&No Amateur&\iibXibLogMpro~(\numiibLogMpro, \numibLogMpro)&\icnoblXicblLogMpro~(\numicblLogMpro, \numicnoblLogMpro)\\
\textbf{PP04}&\textbf{All SN}&\textbf{\iibXibcloseppohfour~(\numiibcloseppohfour, \numibcloseppohfour)}&\textbf{\icnoblXicblcloseppohfour~(\numicblcloseppohfour, \numicnoblcloseppohfour)}\\
&Targeted&\iibXibcloseppohfourtar~(\numiibcloseppohfourtar, \numibcloseppohfourtar)&\icnoblXicblcloseppohfourtar~(\numicblcloseppohfourtar, \numicnoblcloseppohfourtar)\\
&Impartial&\iibXibcloseppohfouruntar~(\numiibcloseppohfouruntar, \numibcloseppohfouruntar)&\icnoblXicblcloseppohfouruntar~(\numicblcloseppohfouruntar, \numicnoblcloseppohfouruntar)\\
&No Amateur&\iibXibcloseppohfourpro~(\numiibcloseppohfourpro, \numibcloseppohfourpro)&\icnoblXicblcloseppohfourpro~(\numicblcloseppohfourpro, \numicnoblcloseppohfourpro)\\
\textbf{T04}&\textbf{All SN}&\textbf{\iibXibclosetohfour~(\numiibclosetohfour, \numibclosetohfour)}&\textbf{\icnoblXicblclosetohfour~(\numicblclosetohfour, \numicnoblclosetohfour)}\\
&Targeted&\iibXibclosetohfourtar~(\numiibclosetohfourtar, \numibclosetohfourtar)&\icnoblXicblclosetohfourtar~(\numicblclosetohfourtar, \numicnoblclosetohfourtar)\\
&Impartial&\iibXibclosetohfouruntar~(\numiibclosetohfouruntar, \numibclosetohfouruntar)&\icnoblXicblclosetohfouruntar~(\numicblclosetohfouruntar, \numicnoblclosetohfouruntar)\\
&No Amateur&\iibXibclosetohfourpro~(\numiibclosetohfourpro, \numibclosetohfourpro)&\icnoblXicblclosetohfourpro~(\numicblclosetohfourpro, \numicnoblclosetohfourpro)\\
\textbf{Host}&\textbf{All SN}&\textbf{\iibXibsnrallhighmass~(\numiibsnrallhighmass, \numibsnrallhighmass)}&\textbf{\icnoblXicblsnrallhighmass~(\numicblsnrallhighmass, \numicnoblsnrallhighmass)}\\
\textbf{Offset}&Targeted&\iibXibsnrallhighmasstar~(\numiibsnrallhighmasstar, \numibsnrallhighmasstar)&\icnoblXicblsnrallhighmasstar~(\numicblsnrallhighmasstar, \numicnoblsnrallhighmasstar)\\
&Impartial&...~(\numiibsnrallhighmassuntar, \numibsnrallhighmassuntar)&\icnoblXicblsnrallhighmassuntar~(\numicblsnrallhighmassuntar, \numicnoblsnrallhighmassuntar)\\
&No Amateur&\iibXibsnrallhighmasspro~(\numiibsnrallhighmasspro, \numibsnrallhighmasspro)&\icnoblXicblsnrallhighmasspro~(\numicblsnrallhighmasspro, \numicnoblsnrallhighmasspro)

\enddata
\tablecomments{P-values and sample sizes from Kolmogorov-Smirnov two-sample comparisons that  include all SN discoveries, targeted SN discoveries, galaxy-impartial SN discoveries, or only professional SN discoveries.
The difference between the metallicity distributions of the hosts of Type Ic-BL and Type Ic SN is statistically even when including only SN discovered by galaxy-impartial hosts. 
The differences between the SN Ib and SN IIb host galaxy $u'$-$z'$ color distributions as well as host galaxy metallicities are statistically significant when including only SN discovered by targeted surveys. 
The statistics from comparing host offsets includes only SN found in massive galaxies (log $M > 9.5$).
}
\label{tab:kspvalues}
\end{deluxetable}

Instead of placing the numerical values of all statistical Kolmogorov-Smirnov (KS) tests in the following descriptions of results, 
we list many of them in Table \ref{tab:kscombo}, which includes comparisons for all types, and Table \ref{tab:kspvalues}, which includes
comparisons for SN IIb and SN Ic-BL, restricted to only targeted and only galaxy-impartial samples. Tables \ref{tab:paptable1} 
and  \ref{tab:paptable2} list the measurements of the SN host galaxies.

\subsection{Host Color and $u'$ Surface Brightness Near Explosion Site}
As can be seen in Figure~\ref{fig:uzcolor}, SN IIb and SN Ic-BL erupt in exceptionally blue environments, while
high $u'$-band surface brightness is typical of SN Ib and SN Ic sites.
SN II sites show substantial overlap in color or surface brightness with the other classes. 
This plot shows $u'$-$z'$ color versus $u'$-band surface brightness, measured inside an aperture consisting of the 20 pixels closest to the 
SN site with $g'$-band S/N $>$ 1. 


The $u'$-$z'$ color near the site of SN 2006aj, the SN-LGRB in our sample, was 0.88 mag.



\subsection{Host Stellar Mass and SN Host Offsets}

Figure~\ref{fig:masses} helps to explain the exceptionally blue $u'$-$z'$ color of SN IIb and SN Ic-BL sites and the high $u'$-band surface
brightnesses near SN Ib and Ic sites. 
At one set of extremes, SN Ic-BL have generally low mass hosts, while SN IIb explode at larger offsets when they occur in galaxies of large masses.
At another extreme, SN Ib and especially SN Ic more often occur inside the $g'$-band half-light radius of massive galaxies, sites expected to have redder color and high surface brightness.




Host galaxy mass is a moderately precise ($\sim$0.1 dex) proxy for chemical abundance (e.g., \citealt{tre04}) which does not suffer from the AGN selection effects. 
The hosts of SN (Ib+Ic), excluding SN Ic-BL, are more massive than SN II hosts. 

The host stellar mass of SN 2006aj, the only SN-LGRB in our sample, was $8.0 \times 10^{10} M_{\odot}$. 
We find with \mbox{$p=\iiXiinsnrall$}~that the SN IIn offset distribution is consistent with the SN II host offset distribution.

\subsection{Oxygen Abundance Measurements Closest to SN Positions}
\label{sec:abundances}
To probe the metallicities of the core-collapse hosts, we measure oxygen abundance from the fiber spectrum with deprojected offset most similar to the SN offset. 
Among the 311 host galaxies with spectroscopic fibers measurements, 139 have multiple SDSS fiber spectra.
SDSS fiber spectra generally target the central regions of host galaxies (see Tables~\ref{tab:selectionImpartial} and \ref{tab:selectionTargeted}), with an average host offset in our sample of \fiboffsetall. 
The low metallicities shown in Figure~\ref{fig:abundances} for SN Ic-BL and SN IIb hosts and high metallicities for SN Ic hosts are consistent with the patterns we see among the species'
colors near explosion sites, host offsets, and host masses. 


\subsubsection{Every Abundance Measurement}

For galaxy-impartial discoveries, SN Ic-BL hosts (\mbox{$n = \numicblcloseppohfouruntar$}) follow a significantly more metal-poor distribution than the hosts of normal SN Ic (\mbox{$n=\numicnoblcloseppohfouruntar$}; \mbox{$p=\icblXicnoblclosetohfouruntar/\icblXicnoblcloseppohfouruntar$}~for T04/PP04 calibrations).
Among the hosts of targeted discoveries, host galaxies of SN IIb (\mbox{$n=\numiibcloseppohfourtar$}) follow a significantly more metal-poor distribution than hosts of SN Ib (\mbox{$n=\numibcloseppohfourtar$}; \mbox{$p=\iibXibclosetohfourtar/\iibXibcloseppohfourtar$}).
Among the hosts of targeted and galaxy-impartial discoveries,
host galaxies of SN IIb (\mbox{$n=\numiibcloseppohfourtar$}) are more metal-poor
than hosts of SN Ib (\mbox{$n=\numibcloseppohfourtar$}; \mbox{$p=\iibXibclosetohfourtar/\iibXibcloseppohfourtar$}). 


The SN II host abundance distribution is more metal-poor than that of the SN Ic hosts, but a selection effect may inflate any difference.  
A higher fraction of SN II (\sniiagn) than SN (Ib+Ic) host galaxy fiber spectra (\snibcagn) have the emission line ratios of AGN (see Tables~\ref{tab:selectionImpartial} and \ref{tab:selectionTargeted}), which makes spectra unusable for abundance analysis. AGN occur primarily in massive, metal-rich galaxies ($M>10^{10} M_{\odot}$; \citealt{kauffmann03agn}), so rejecting AGN spectra removes a higher fraction of metal-rich SN II hosts than of SN Ib/c hosts.
We note that the presence of nuclear activity in a host galaxy does not necessarily mean that every SDSS host spectrum will show contamination, because
SDSS fibers are sometimes offset from the nucleus (see Tables~\ref{tab:selectionImpartial} and \ref{tab:selectionTargeted}).
A host galaxy with mass $10^{10.5} M_{\odot}$, typical of an AGN host, will have an oxygen abundance of $\sim$9 dex (T04) and $\sim$8.75 dex (PP04) (e.g., \citealt{tre04}). 

SN IIn hosts follow a similar distribution to that of the entire SN II sample (\mbox{$p=\iinXiiclosetohfour/\iinXiicloseppohfour$}).



\subsubsection{When Fiber and SN Host Offsets Are Similar}
\label{sec:simoffsets}

Most galaxies have metallicity gradients, with abundance
declining away from the galaxy center.
Van Zee et al. \citeyear{vanzee98} found, for example, a mean radial abundance gradient of -0.052 dex kpc$^{-1}$ for a sample of 11 NGC host galaxies.
To assemble improved proxies for metallicity at the SN location, we selected fibers whose deprojected host offset (away from the galaxy center) was within 3 kpc of that of a SN.

Among these fibers, the SN Ic-BL host spectra are significantly more metal-poor than 
both the SN Ib and SN Ic spectra.
Without making a correction for the difference between fractions of SN II and SN Ib/c host SDSS spectra with no abundance estimate due to AGN contamination, 
the SN II host fibers (with similar host offset) are significantly less metal-rich than that of SN Ic host fibers. 


Median offset differences between the SDSS fiber and SN location (in kpc): SN Ib (1.02), SN Ic (1.32), SN Ic-BL (1.56), SN II (1.11), SN IIb (1.66).

\subsection{Host Specific Star Formation Rate from Fiber Spectra}

SDSS spectra provide an estimate of the specific SFR (SFR M$_{\odot}^{-1}$ yr$^{-1}$) within the aperture of the fiber, which generally targets the host galaxy within the $g'$-band half-light radius. 
As can be seen in Figure~\ref{fig:ssfr}, there is a progression of increasing specific SFR from SN II to SN Ib to SN Ic host spectra. 
SN Ic-BL host spectra also have significantly greater specific SFR than SN II host spectra. 

Using visual inspection, we identified fibers that target the host galaxy nucleus to $z=0.04$ where the 3$"$ fiber aperture primarily samples nuclear light. 
These spectra yield significant evidence that the nuclei 
of SN (Ib+Ic) host galaxies have greater specific star formation rates than those of SN II host galaxies (\mbox{$p=\ibicXiiSSFRnuc$}).
Strong central star formation among SN (Ib+Ic) hosts may overwhelm AGN-patterned emission and explain the relatively low AGN fraction among SN (Ib+Ic) host galaxies.


\subsection{Extinction Inferred from Spectra}




Although SN Ic hosts have stronger specific SFR 
within the half-light radius, the region where most SN Ic explode, 
the sites of SN Ic are not bluer than those of SN II (see Figure~\ref{fig:uzcolor}).
Figure~\ref{fig:abextinct} shows that the high extinction of SN (Ib+Ic) host galaxies, measured from host spectra, may 
redden ongoing star formation in SN Ic host galaxies. 
 
The host galaxies of SN IIb have less extinction than SN (Ib+Ic) host galaxies. 
The average extinction difference between SN (Ib+Ic) and SN IIb hosts is $A_V\approx0.5$ mag, a $u'$-$z'$  reddening of $\sim$0.6 mag. 
The approximately similar internal extinctions of SN II and SN IIb hosts, however, suggest that the stellar populations near SN IIb likely are intrinsically bluer than those near SN II sites. 


 SN (Ib+Ic) host reddening is consistent (\mbox{$p=\droutibic$}) with that estimated along the line-of-sight to 19 SN (Ib+Ic) from their light curve colors by \citet{drout11} using an empirical model of SN Ib/c photometric color evolution. 
Comparison between the \citet{drout11} sample and the SN II host $A_{V}$ distribution yields \mbox{$p=\droutii$}.
Here we plot only the \citet{drout11} Gold and Silver SN.  There is a median $A_{V}\approx1.2$ mag extinction through SN (Ib+Ic) host fiber apertures ($E(B-V)\approx0.4$ mag).

\begin{figure*}
\begin{center}
\includegraphics[angle=0,width=6.0in]{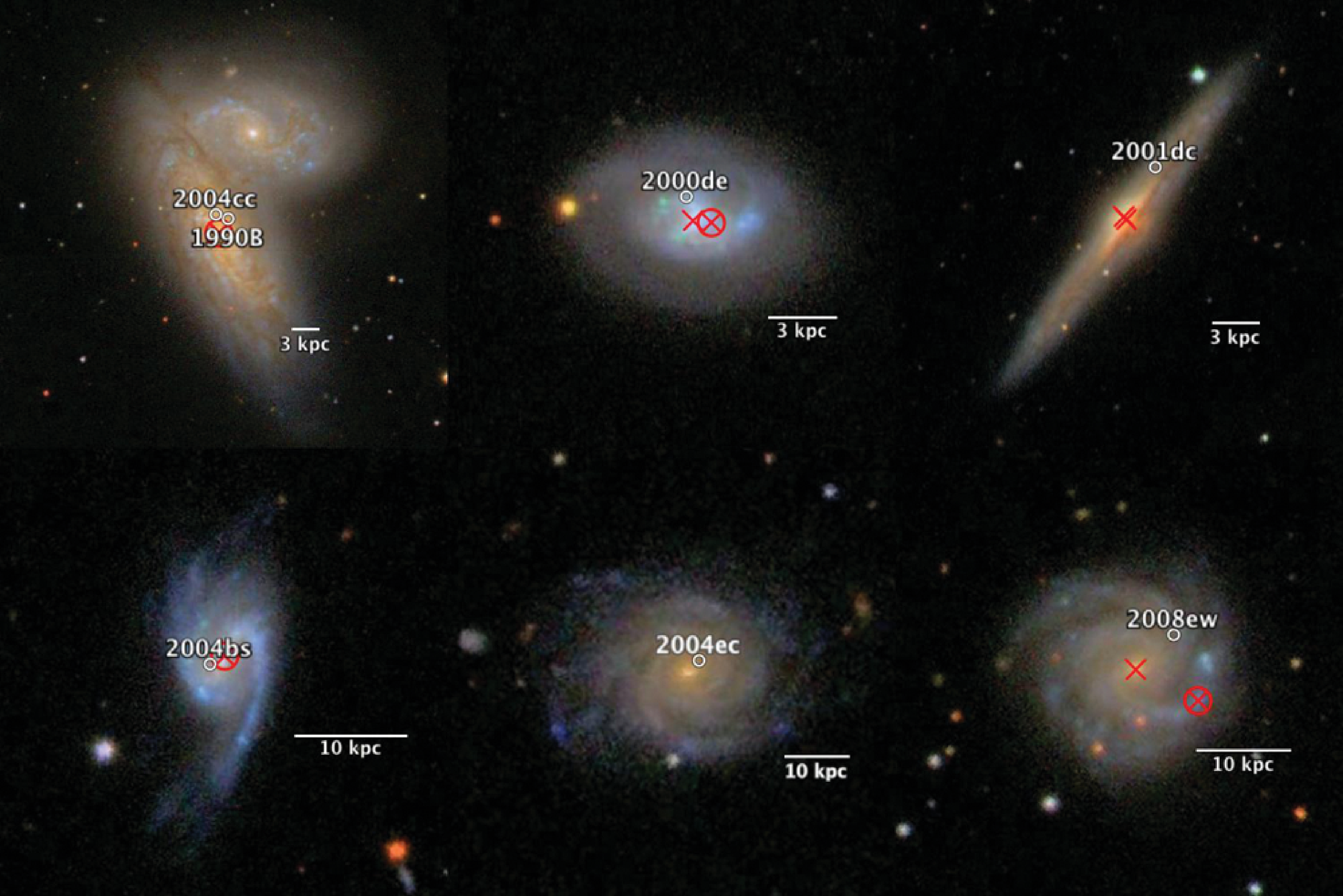}
\end{center}
\caption{SDSS color composite images of 6 SN Ib, SN Ic, and SN II host galaxies in our sample. These include: SN 1990B (Ic), SN 2004cc (Ic), SN 2000de (Ib), SN 2001dc (IIP), SN 2004bs (Ib), SN 2004ec (IIn), and SN 2008ew (Ic). 
Red cross hatches show SDSS fiber positions yielding 
oxygen abundance measurements. An additional red circle marks fibers whose host offsets are within 3 kpc of the SN offset.
 }
\label{fig:hostcolorsibicii}
\end{figure*}

\begin{figure*}
\begin{center}
\includegraphics[angle=0,width=6.0in]{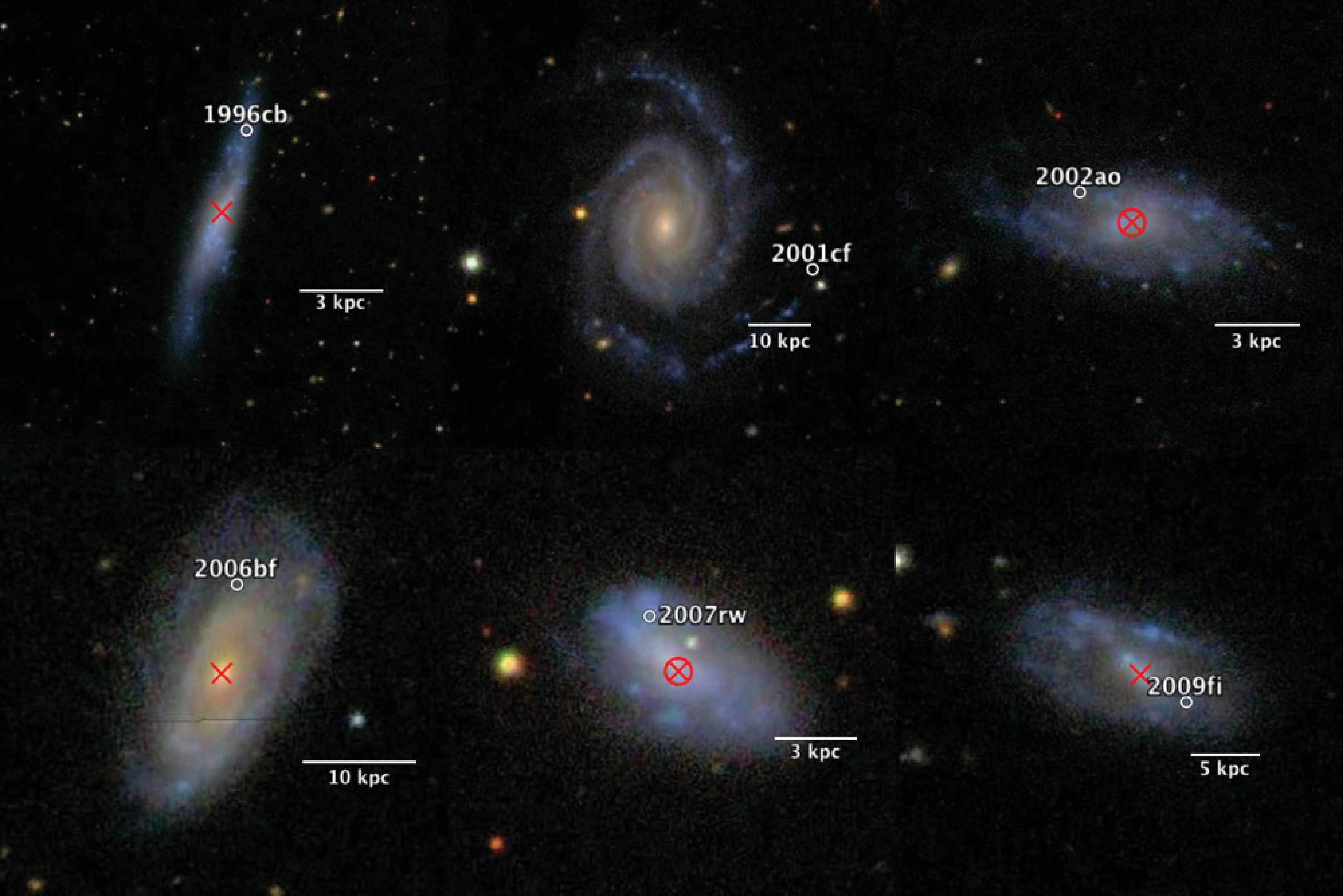}
\end{center}
\caption{SDSS color composite images of 6 SN IIb in our sample. 
Their local environments are
substantially bluer than those of SN Ib, SN Ic, and SN II. 
Red cross hatches show SDSS fiber positions yielding 
oxygen abundance measurements. An additional red circle marks fibers whose host offsets are within 3 kpc of the SN offset.
}
\label{fig:hostcolorsiib}
\end{figure*}

\begin{figure*}
\begin{center}
\includegraphics[angle=0,width=6.0in]{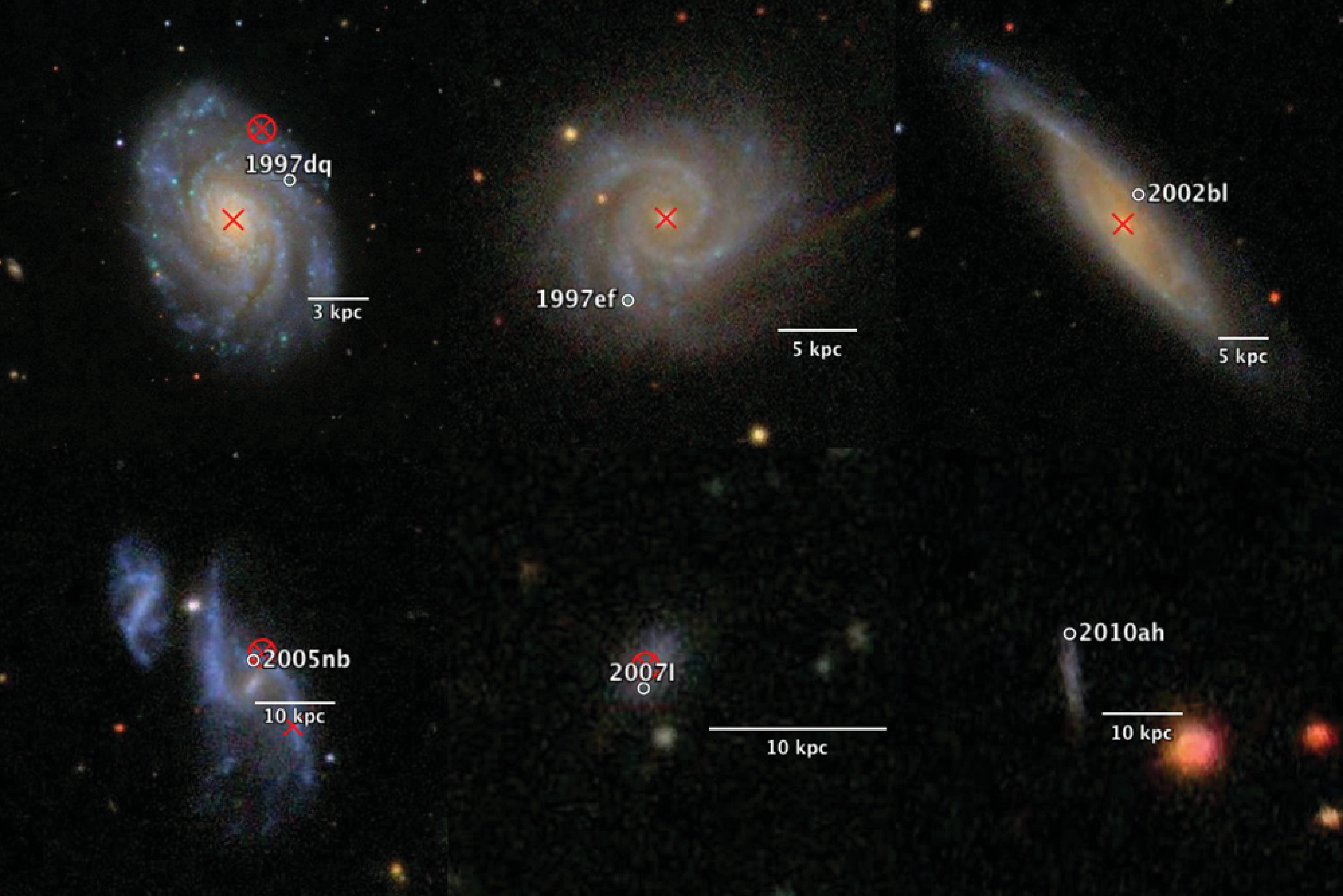}
\end{center}
\caption{SDSS color composite images of 6 SN Ic-BL in our sample. 
SN Ic-BL in our sample occurred preferentially in lower-mass, low-metallicity host galaxies.
Red cross hatches show SDSS fiber positions yielding 
oxygen abundance measurements. An additional red circle marks fibers whose host offsets are within 3 kpc of the SN offset.
 }
\label{fig:hostcolorsicbl}
\end{figure*}

\section{Relative Frequencies of Core-Collapse SN as a Function of Metallicity}
\label{sec:rates}

We plot the ratio of stripped-envelope SN (including SN IIb) to SN II in our sample with increasing host galaxy oxygen abundance in Figure~\ref{fig:ratios}. 
Vertical error bars show the Poisson uncertainties, while horizontal bars indicate the range of metallicities in each bin. 
The color points are calculated from successful \textit{spectroscopic} metallicity measurements, while the gray points are estimated using \textit{stellar mass} as a metallicity proxy, applying the \citet{tre04} mass-metallicity relation. 

AGN emission, present disproportionately in SN II host spectra, is found primarily in high-mass, high-metallicity galaxies. This selection effect misleadingly \textit{inflates} the apparent ratio \mbox{SN (Ib+Ic) / SN II} (color points) in the highest metallicity bin. Indeed, the ratio at high metallicity calculated instead using stellar masses as a proxy (which has no similar selection effect) is significantly
lower (gray points). Earlier efforts using SDSS fiber spectra (i.e., \citealt{pri08}), which also exclude AGN-contaminated spectra, have not noted this strong selection effect at high metallicity. 

While we have attempted to identify and limit systematic effects, 
we cannot fully know the selection biases (e.g., from classification, detection) that may affect the measured ratios.
We compare the relative rates of SN types in our sample to the model predictions for single, rotating progenitors \citep{georgy09}, single, non-rotating progenitors \citep{eld08}, and binary  progenitors \citep{eld08}. Plotted \citet{georgy09} predictions were made with the assumption that a minimum helium envelope of 0.6 M$_{\odot}$ separates the progenitors of SN Ib and SN Ic. Because core collapse to a black hole may not yield a SN explosion (e.g., \citealt{fry99}), especially if high angular momentum does not support an accretion disk (\citealt{wo93}), \citet{georgy09} calculated predictions where viable SN occur after core collapse to (a) only neutron stars and (b) neutron stars and black holes. These predictions adopted 2.7 M$_{\odot}$ as the maximum mass of neutron star (\citealt{shapiro83}) and use the \citet{hirs05} relation between neutron star mass and the mass of the carbon-oxygen core.

Model predictions are parameterized by $Z/Z_{\odot}$, requiring us to subtract the solar value from 12+log(O/H) estimates for each host galaxy to compute log($Z/Z_{\odot}$).
The value of the solar metallicity is, however, not well constrained.
Atmospheric modeling favors lower solar values (e.g., 12+log(O/H)=8.69; \citealt{asplund09}) than 
helioseismic analyses (e.g., 12+log(O/H)=8.86; \citealt{delahaye10}).  Here we use the helioseismic value of \citet{delahaye10}.

Returning to our data, we note that the spectroscopic oxygen abundance measurements should, on average, be \textit{overestimates} of the oxygen abundance at the SN site because the SDSS fibers are concentrated toward the inner regions of the galaxies.
Likewise, abundances calculated from host masses and the \citet{tre04} $M-Z$ relation should also 
be overestimates, because the \citet{tre04} relation is a fit to SDSS stellar masses and fiber metallicities. 



Wind-driven mass losses by single stars at low metallicity are thought to be comparatively modest (e.g., \citealt{eld08}; \citealt{smar09}; \citealt{yoon10}).
To explain the presence of stripped-envelope SN in low-metallicity environments in Figure~\ref{fig:ratios}, model comparison requires that collapse to a black hole or binary stripping be a viable route to stripped-envelope explosions. 
Single star models predict that the only stars that lose their outer envelopes at low metallicity are sufficiently massive that they collapse to black holes
(e.g., \citealt{eld08}).


\citet{smith11frac} note that single star models use constant rates of wind-driven mass
loss substantially greater than those observed, although episodic mass loss may speed loss of the outer envelopes.
Lower wind-loss rates would imply a diminished fraction of single progenitors. 

Splitting our sample in two at $z = 0.015$, the same trends persist in both the low and high redshift subsamples, providing some evidence that 
they do not result from luminosity-dependent selection effects.

\section{Tests and Potential Systematic Effects}
\label{sec:tests}

\subsection{Fiber Aperture Coverage}

SDSS spectroscopic fiber apertures have a fixed radius of 1.5$"$. 
At increasing redshift, this aperture radius corresponds to larger physical scales: 0.3 kpc at $z=0.01$; 0.6 kpc at $z=0.02$; and
1.17 kpc at $z=0.04$. 
For a sample of 11 NGC spiral galaxies, \citet{vanzee98} found a mean radial abundance gradient of -0.052 dex kpc$^{-1}$.
While metallicity gradients vary among galaxies and have some dependence on, for example, host galaxy morphology (e.g., \citealt{kew06}) and the metallicity calibration (e.g., \citealt{moose10}), we use this as a representative value. 

Within the targeted sample ($z < 0.023$), nuclear spectroscopic fibers extend at most 0.68 kpc away from the host center,
corresponding to systematic shifts of order only $\sim$0.025 dex.
Galaxy-impartial SN discoveries (to $z < 0.08$) account for a significant fraction of only the SN Ic-BL sample. 
The difference between the median abundances for SN Ic-BL and SN Ic hosts is $\sim$0.5 dex, substantially greater than an aperture effect may yield.



\subsection{Classification}

There may be variation among the classification practices 
of the different surveys that contribute to our samples. A
concern is that surveys that monitor different host
galaxy populations (e.g., galaxy-impartial and targeted)  could have
different classification practices, such as use of automated
classification tools (e.g., SNID) or multi-epoch spectroscopic 
follow up. For instance, the helium lines that identify SN Ib
often emerge only after a couple of weeks (e.g., \citealt{li11a}).




\subsection{Fiber Targeting}

The SDSS object detection algorithm mistakenly split many galaxies of large angular size into two or more components [see Fig. 9 of \citet{bl05}]. 
The SDSS targeting algorithm then placed fibers on these
false components, sometimes at significant offset from the true galaxy center. 
The error rate of these algorithms could depend on galaxy morphology (e.g., irregularity or an interacting neighbor), and
we checked whether the offsets of fiber measurements depend on SN type. However, we found no evidence of strong variation with SN type.

SDSS fibers often target the local maxima of galaxy light distributions, including host nuclei and bright HII regions.
In our sample, fibers have mean offset of \fiboffsetall, while matched fibers  (where \mbox{$|r_{env} - r_{fiber}| < $ 3 kpc}) have mean offset of $0.55 \times r_{half-light}$.
Therefore, fiber sites are highly likely to be more metal-rich on average than they would be if SDSS fibers sampled galaxy light distributions more democratically. 
However, the fibers' offset distribution does not vary strongly with SN type. 

The lifetimes of HII regions may be shorter than the those of the progenitors (private communication; N. Smith), and the signal at the SN site may be too weak.
Programs that take host spectra at the location of the SN (e.g., \citealt{anderson10}; \citealt{mod11}; \citealt{leloudas11})
may only extract a metallicity estimate when there is sufficient nebular emission through the slit. 
Any such S/N requirement could possibly act as a type-dependent selection effect. 
The SDSS targets bright nuclei or HII regions, moderating any such effect in our analysis. 

\section{Discussion}
\label{sec:discussion}

Our study of the host environments of core-collapse SN reported to the IAU or discovered by the PTF has revealed several statistically significant patterns.
While we have constructed the sample to limit the influence of potential selection effects,
we have a limited knowledge of the contributing SN search programs (e.g., cadence, limiting magnitude, classification methods).
We have found that the $u'$-$z'$ colors of the SN IIb and SN Ic-BL (without an associated LGRB) environments are blue in comparison to those of other stripped-envelope SN environments (see representative images in Figures \ref{fig:hostcolorsibicii}, \ref{fig:hostcolorsiib}, and \ref{fig:hostcolorsicbl}.
The host specific SFR (SFR M$_{\odot}^{-1}$ yr$^{-1}$) is higher, on average, for types whose SN spectra indicate more complete loss of the progenitor's outer envelopes (i.e., SN Ib, SN Ic, SN Ic-BL).
Spectroscopy also shows that, in our sample, SN Ic-BL host galaxies are more metal-poor than the hosts of normal SN Ic explosions, while
SN IIb hosts also are more metal-poor (for one of two abundance diagnostics) and have less extinction, on average, than SN Ib or SN Ic host galaxies.


A surprising effect is that spectroscopic contamination by AGN is higher among SN II hosts than SN (Ib+Ic) hosts. This is important to the correct interpretation of host galaxy properties from SDSS spectroscopy.


 

This study is statistical, and we have shown only that samples are drawn from differing underlying distributions in our comparisons.
The distinctions we present are consistent with even considerable variation among the environments of individual examples of each SN type.

\subsection{Synthesizing Patterns}

There are strong connections among the type-dependent patterns in host galaxy photometry and spectroscopy:

\begin{itemize}

\item 
Host galaxies of SN Ic-BL in the sample generally have low mass and high specific SFR, helping to explain the blue colors
at broad-lined SN Ic explosion sites. 
The SN IIb typically are found beyond the $g'$-band half-light radius in massive hosts,
offering explanation for the blue colors of their sites.
The SN Ic-BL and SN IIb host galaxy fiber spectra have lower abundances than SN Ic and SN Ib host galaxy fiber spectra closest in radius
to the explosion site, respectively,
although the SN IIb-SN Ib difference is significant for only one of two strong-line diagnostics.

\item SN Ic often erupt at small offsets in massive galaxies with strong specific SFR, high oxygen abundance, and high extinction measured from fiber spectra. These
fibers generally collect light from within the host's
$g'$-band half-light radius.  
These explosion sites help to explain the high surface brightnesses near SN Ic explosion sites.
High interstellar reddening helps to explain why the
colors near SN Ic sites have colors similar to those of SN II, despite their hosts' high specific SFR. 



\end{itemize}

The $u'$-$z'$ color and $u'$ surface brightness near SN explosion sites 
considerably  separate SN IIb, SN Ic, and SN Ic (see Figure \ref{fig:uzcolor}). SN Ic sites predominantly have high surface brightness, 
while SN IIb populate lower-surface brightness but extremely blue environments. SN Ib 
largely occupy the parameter space between the SN IIb and the SN Ic. 
By contrast, however, the explosion sites of SN II have no specific locus in the color-brightness plane.

These patterns suggest that the fraction of the overall stripped-envelope SN (Ib+Ic+IIb) to Type II SN may not vary strongly with environment. 
Perhaps the stars that may explode as one stripped-envelope species at 
one value of mass in one environment instead exploded as another stripped-envelope species
in other environments where, for example, the chemistry is different.


\subsection{SN Ib, SN IIb, and SN II Environments}

The best-studied example of a Type IIb, SN 1993J, exploded at a distance of only 3.6 Mpc in M81 (\citealt{fil93}; \citealt{math00}), and archival imaging revealed that its progenitor was a K-type supergiant (\citealt{aldering94}). HST imaging after the SN disappeared found evidence for a B-type supergiant binary companion (\citealt{vd02}; \citealt{maun04}). More recent studies of the sites of other SN IIb suggest, however, that a fraction of the SN IIb population may erupt from massive single stars (e.g., \citealt{crockett08}). \citet{chev10} analyzed the radio emission, optical shock breakout, and nebular emission of a sample of SN IIb to constrain the extent of their progenitors' envelopes and the properties of their circumstellar material. They favor two progenitor populations: (a) extended progenitors (SN 1993J, SN 2001gd) with hydrogen envelope mass greater than $\sim 0.1 M_{\odot}$ and slow, dense winds and (b) more compact and massive Wolf-Rayet progenitors (SN 1996cb, SN 2001ig, SN 2003bg, SN 2008ax, and SN 2008bo) with a less massive hydrogen envelope and lower density winds. PTF11eon/SN 2011dh, a SN IIb (\citealt{arcavi11}; \citealt{marion11}), was recently discovered by amateur astronomer Amadee Riou in M51, where pre-explosion HST imaging exists of the SN site. Analysis of the archival images finds evidence for a supergiant with $T_{eff} \approx 6000$ K at or near the SN site (\citealt{vandyk11dh}; \citealt{maund11dh}). Radio and X-ray observations (\citealt{soderberg12}) and the optical spectroscopic and photometric evolution (\citealt{arcavi11}; Marion et al. 2012, in preparation) both favor a compact progenitor, however, suggesting that this star may be a binary companion or not associated with the SN. 

Our analysis finds three statistically significant, plausibly related patterns in the host environments of SN IIb:
SN IIb environments are bluer than the environments of SN Ib, SN Ic, and SN II;
their explosion sites may be more metal-poor than those of SN Ib or SN Ic (significant for one of two abundance diagnostics); and 
their host galaxy interstellar extinction is less than that of SN (Ib+Ic).
These trends are statistically significant even when we analyze only the locations of 
targeted SN. 

An unambiguous implication of the exceptionally blue colors of SN IIb environments is that the Type IIb 
progenitor population is distinct from that of Type Ib explosions. Lack of hydrogen features near
maximum light in SN Ib spectra may reflect a more extensive loss of the progenitor's hydrogen envelope.
Comparatively metal-poor SN IIb host galaxies suggest that metals may play an important role in
achieving this loss of the outer envelope.

The SN IIb population may erupt from a combination of massive single stars and progenitors in close binary systems, so a possibility is that the blue colors of SN IIb environments indicate higher binary fractions. Although the current examples of each class are few, future efforts 
may be able to draw distinctions between the environments of the compact and extended SN IIb progenitors proposed by \citet{chev10}.

\citet{li11b} recently reported that the hosts of SN IIb detected by LOSS had greater $K$-band luminosities than SN II-P hosts (with \mbox{$p=6.9$}\%). 
Lower SN IIb progenitor metallicities are consistent with the PTF's diminished fraction of SN IIb and SN Ic-BL in `giant,' presumably metal-rich galaxies, than in `dwarf' galaxies: 1 SN Ib, 3 SN IIb, and 2 SN Ic-BL, and 9 SN II in `dwarf' galaxies, and 2 SN Ib, 2 SN IIb, 7 SN Ic, 1 SN Ic-BL and 42 SN II in `giant' galaxies \citep{arcavi10}. 

\subsection{SN Ic-BL Environments}

Type Ic-BL are the SN that have been associated with coincident LGRB explosions (\citealt{ga98}; \citealt{ma03}; \citealt{st03}; \citealt{hj03}; \citealt{malesani04}; \citealt{mod06}; \citealt{sanders11}; see \citealt{woosley06} and \citealt{modjaz11rev} for reviews). \citet{mod08} showed that SN Ic-BL with associated LGRB prefer more metal-poor environments than do SN Ic-BL without an LGRB (but see \citealt{lev10highmet}). 

We find that host galaxies of SN Ic-BL (without an associated LGRB) follow a significantly more metal-poor distribution than the hosts of normal SN Ic (or SN Ib) explosions, even when only galaxy-impartial discoveries are considered. 
The colors of SN Ic-BL local environments also follow a bluer distribution than those of SN Ic, further evidence for different progenitor populations. SN Ic-BL host galaxies have strong specific SFRs, similar to those of normal SN Ic.


Lower Type Ic-BL progenitor oxygen abundances may imply reduced rates
of wind-driven mass loss, potentially enabling SN Ic-BL progenitor to retain greater angular momentum (e.g., \citealt{kud02}; \citealt{heg03}; \citealt{eld04}; \citealt{vin05}). 
High angular momentum before the explosion may be important to the production of high velocity ejecta (\citealt{wo93}; \citealt{thompson04}). 
Nonetheless, the means by which SN Ic-BL progenitors shed their outer envelopes, if not through their high metallicity, 
needs explanation and may involve Roche lobe overflow (\citealt{pods92}; \citealt{nomo95}), stellar mergers (\citealt{pod10}), or perhaps deep mixing.

Here our measurements support a picture where both SN Ib and SN Ic have more metal-rich hosts on average than SN Ic-BL,
consistent with the host galaxy magnitudes measured by \citet{arcavi10}.  It presents a contrast with the results of \citet{mod11} who recently measured the oxygen abundances at the \textit{sites} of SN Ic-BL, SN (Ib+IIb), and SN Ic.
There the SN Ic-BL distribution falls intermediate between those of SN (Ib+IIb) and SN Ic, although it is more similar to the comparatively metal-poor SN (Ib+IIb) distribution and neither comparison is statistically significant. These contrasting trends may relate to fact that \citet{mod11} constructed their samples for each SN type from approximately equal numbers of 
galaxy-impartial and targeted SN discoveries, or the inclusion of SN IIb (which we find inhabit metal-poor environments) with SN Ib.
\citet{mod11} measurements were also taken at the explosion site, which may often differ significantly from the 
host abundance measured from SDSS fiber spectra (0.13 dex average disagreement with \textit{nuclear} fiber measurements).

\citet{svensson10} found that host galaxies of LGRBs had smaller star masses than core-collapse SN hosts and had high surface brightness and more massive stellar populations. The only SN-LGRB that met our sample criteria, SN 2006aj, has low host stellar mass and comparatively blue $u'$-$z'$ color near the explosion site. 

\subsection{SN Ib, SN Ic, and SN II Environments}

In an earlier paper \citep{kel08}, we showed that, while the positions of the other core-collapse SN follow the distribution of their hosts' light, Type Ic SN trace the brightest regions of their host galaxies in a pattern similar to that followed by LGRB (\citealt{fr06}). Possible explanations for this pattern include shorter lifetimes and higher masses of SN Ic progenitors (\citealt{raskin08}; \citealt{leloudas10}; \citealt{eldridge11}) as well as preference for metal-rich regions near the centers of hosts.  
\citet{and08} showed, subsequently, that SN Ic also track their hosts' H$\alpha$ emission more closely than SN II (their comparison with SN Ib lacked statistical significance).

The SDSS fiber spectra of core-collapse host galaxies, which generally sample inside of the $g'$-band 
half-light radius, reveal an increasing progression of specific SFR from SN II to SN Ib to SN Ic (and SN Ic-BL) hosts.
This pattern persists when we study only the spectra from fibers targeting the host nucleus.
SN Ic explode at comparatively small host offset, linking them to the strong star formation near their hosts' centers. 


We find that the central star formation that yields SN Ic generally has high chemical abundance and extinction from interstellar dust.
A SN Ic progenitor population tracking high metallicity would be expected to explode in massive galaxies with  strong star formation 
in metal-rich gas near their centers, the pattern we observe.

The colors of SN Ib and SN Ic explosion sites may offer evidence that their progenitors 
are also younger and more massive than the progenitors of SN II.
The distribution of the apparent  $u'$-$z'$ color at SN Ib and SN Ic sites is similar
to that at SN II sites. However, we find that SN (Ib+Ic) host galaxies have
higher interstellar extinction ($\Delta A_{V} \approx 0.5$ mag). This suggests that SN (Ib+Ic) sites have intrinsically 
bluer color than SN II sites, perhaps indicative of younger progenitor stellar populations.

 



SN Ib explosion sites have higher $u'$-band surface brightnesses than SN II sites, while
SN Ib host galaxies generally have lower abundance than SN Ic in our sample. 
There is no statistically significant difference between the SN Ib and SN Ic host offset distributions in our sample (\mbox{$p=\ibXicnoblsnrall$}), which may imply that 
host offset cannot, on its own, explain the uniquely strong association of SN Ic with bright host galaxy pixels (\citealt{kel08}).




While analyses of pre- and post-explosion imaging have not yet identified a progenitor of a SN Ib or SN Ic, red supergiants have been found  
at the sites of SN II-P explosions (e.g., \citealt{barth96}; \citealt{vd99}, \citeyear{vd03a}, \citeyear{vd03b}, \citeyear{vandyk2010}; \citealt{smartt01}, \citeyear{smartt03}, \citeyear{smartt04}; \citealt{li05}, \citeyear{li07}; \citealt{maund05b}). \citet{smartt09} favor a 8.5-16.5 $M_{\odot}$ mass range for SN II progenitors, although 
extinction along the line of sight to the progenitors is not well constrained (e.g., \citealt{walmswell11}). 
\citet{smith11frac} note that Wolf Rayet stars in binary systems, possible progenitors of SN Ib and SN Ic, 
are expected to be less luminous than single Wolf Rayet stars. 
Brighter  companions may outshine Wolf Rayet progenitors,
although mass-gaining companions may, in some cases,  explode first (\citealt{pods92}; \citealt{eldridge11}). 
Even for progenitors with close binary companions, 
metallicity and mass are expected to be important in determining the
composition of the outer envelope, even though substantial mass loss may occur through Roche lobe overflow (\citealt{smith11frac}; \citealt{yoon10}; \citealt{eldridge11}).

\citet{pran03} and \citet{boi09} found that SN (Ib+Ib/c+Ic) hosts have greater absolute $M_{B}$ luminosities than SN II hosts.
\citet{pri08} presented the first comparison between the oxygen abundances of SN (Ib+Ic) and SN II host galaxies 
with large sample sizes. 
Using the T04 metallicities available for the SDSS DR4 spectra, 
they found \mbox{$p=5$}\% evidence for a difference (the sample may not have been large enough to determine the effect of AGN contamination). 
Van den Bergh \citeyear{vanden97}, \citet{tsvetkov04}, \citet{hakobyan09}, \citet{and09}, and \citet{leaman11} have found that SN (Ib+Ib/c+Ic) occur preferentially toward galaxy centers, where oxygen abundances are generally higher. 
\citet{habergham10}, examining 178 host galaxies for evidence of interaction, and \citet{anderson11}, in a study of SN sites in Arp 299, have explored explanations for these patterns.

\citet{mod11} find a significant difference ($\sim$0.2 dex on average) between the oxygen abundances at the sites of 12 SN Ic and a mixed sample of 16 SN (Ib+IIb) for one of three oxygen abundance calibrations (although see \citet{anderson10} and \citet{leloudas11}). 
When only abundances measured at the SN site from these three studies are compared, a significant difference between SN Ib and SN Ic metallicities computed with the \citet{pettini04} diagnostic is evident (M. Modjaz, private comm. and in preparation). 


\subsection{SN IIn Environments}

Among our set of host measurements, we find no statistically significant differences between the characteristics of SN IIn host environments and those of normal SN II. 
\citet{and09}, who have also studied SN IIn explosion sites, found no significant difference between the mean radial offsets of 12 SN IIn and 35 SN IIP from the host galaxy center.

Narrow line emission characterizes SN IIn spectra (\citealt{sch90}) and is thought to be the result of the interaction of the ejecta with high density surrounding material. 
The existence of dense circumstellar material likely indicates strong pre-explosion mass loss (e.g., \citealt{chugai94}) and
can increase the optical luminosity of the SN by thermalizing the emerging blast wave (e.g., \citealt{woosley07}; \citealt{smith11frac}; \citealt{vanmarle10}).

Luminous Blue Variable (LBV) stars (e.g., $\eta$ Car), with their high mass loss rates  ($>10^{-4}M_{\odot}$ yr$^{-1}$), have been suggested as candidate progenitors, although standard stellar modeling positions the LBV period before an ultimate Wolf-Rayet phase (e.g., \citealt{langer93}; \citealt{maeder05}).
\citet{dwark11} has recently suggested that observations may only present a convincing case for an LBV progenitor in the case of SN 2005gl (\citealt{galyam07}; \citealt{galyam09}).
Other means of potentially producing regions of high density circumstellar material include, for example, pulsation-driven superwinds from red supergiants (RSGs) (\citealt{yooncant10}).

\section{Conclusions}
\label{sec:conclusion}

We have analyzed the properties of the environments of nearby SN reported by
targeted and galaxy-impartial searches. 
Most of our data come from a small number of well-defined surveys, but we have included supernovae discovered by many individuals and groups.  It is not possible for us to characterize the systematic effects introduced by each search.  However, we have characterized the searches by their fundamental techniques and applied reasonable redshift limits to limit the strength of possible bias.  We show that these details do not dominate the patterns we find.


The SN IIb and SN Ic-BL in our sample erupt in environments with exceptionally blue color. SN IIb sites often have large host offsets, while SN Ic-BL generally have comparatively low mass host galaxies.
By contrast, SN Ib and especially SN Ic environments have less extreme colors, similar to those of SN II sites, but with exceptionally high $u'$-band surface brightness. SN Ib and SN Ic generally erupt from regions within the $g'$-band half-light radii
of high stellar mass galaxies. The  colors and surface brightnesses of
SN II as well as SN IIn environments show no strong distinguishing pattern.

The centers of SN Ic host galaxies are generally dusty, metal-rich, and have high specific SFR.
Stronger interstellar extinction associated with SN Ic sites may explain why they are not bluer than SN II sites, 
despite higher specific SFR. 
The central regions of SN Ib host galaxies are less metal-rich and have smaller specific 
SFR than those of SN Ic hosts.



We find that the SN IIb host galaxy spectra closest in radius to the explosion site in our sample
are more metal poor than the SN Ib host galaxy spectra, although this difference is statistically significant for only one of two strong-line diagnostics. 
SN Ic-BL host galaxies are also less metal-rich than SN Ic host galaxies, even among only
galaxy-impartial discoveries. 

The specific SFR measured from fiber spectra is higher, on average, for types whose SN spectra indicate more complete loss of the progenitor's outer envelopes (e.g., SN Ic, SN Ic-BL).
Even among only spectra of galaxy nuclei, SN (Ib+Ic) host spectra have stronger specific SFR than SN II host spectra.

The non-negligible fraction of stripped-envelope SN in low-metallicity host galaxies 
may indicate that some stripped-envelope SN have binary progenitors or, alternatively, single progenitors that collapse to a black hole.


\citet{drout11} have estimated the line-of-sight extinction instead inferred from the colors of SN light curves.
The interstellar reddening we find from SDSS fiber spectra of SN Ib and SN Ic hosts yield consistent values of $A_V$,
although the SDSS fibers are generally positioned away from the explosion site.

AGN emission, which makes spectra unusable for abundance measurements and is found primarily in high-metallicity galaxies, leads us to exclude a 
larger fraction of SN II (\sniiagn) than SN (Ib+Ic) host spectra (\snibcagn). 
This produces an overestimate of \mbox{SN (Ib+Ic) / SN II} in high-metallicity environments from SDSS spectra alone.  
The ratio is lower when we use host stellar mass as an oxygen abundance proxy, impervious to AGN. 

Stellar mass estimates, robust to AGN contamination, provide evidence that \mbox{SN (Ib+Ic) / SN II} increases in more massive, metal-rich galaxies,
a trend that retains significance when we consider only targeted SN discoveries.

None of the host measurements reveals a strong difference between
SN IIn and normal SN II explosion environments.

The accelerating rate of SN discovery promises to yield, over the next decade, 
much larger samples of each of the core-collapse species that we study in
this paper.
We urge the public archiving of spectra so that analyses can assign SN to consistent spectroscopic
classes.  Future study of explosion sites, aided by improved
position information and uniform classification, will be a powerful tool to study progenitor properties
and the evolution of massive stars.

\acknowledgements

Thanks especially to Maryam Modjaz for her perceptive comments as well as revised spectroscopic classifications and to David Burke for help with both supporting observations and editorial feedback. We also thank Peter Challis, Howie Marion, Nadia Zakamska, Georgios Leloudas, Shizuka Akiyama, Steve Allen, Roger Romani, Sung-Chul Yoon, Michael Blanton, Nathan Smith, Anja von der Linden, Mark Allen, and Douglas Applegate for their advice and help. We acknowledge the MPA-JHU collaboration for making their catalog publicly available and Google Sky for help in producing color galaxy images. RPK's supernova research at the Center for Astrophysics is supported by NSF grant AST0907903.

Funding for the SDSS and SDSS-II has been provided by the Alfred P. Sloan Foundation, the Participating Institutions, the National Science Foundation, the U.S. Department of Energy, the National Aeronautics and Space Administration, the Japanese Monbukagakusho, the Max Planck Society, and the Higher Education Funding Council for England. 

The SDSS is managed by the Astrophysical Research Consortium for the Participating Institutions. The Participating Institutions are the American Museum of Natural History, Astrophysical Institute Potsdam, University of Basel, Cambridge University, Case Western Reserve University, University of Chicago, Drexel University, Fermilab, the Institute for Advanced Study, the Japan Participation Group, Johns Hopkins University, the Joint Institute for Nuclear Astrophysics, the Kavli Institute for Particle Astrophysics and Cosmology, the Korean Scientist Group, the Chinese Academy of Sciences (LAMOST), Los Alamos National Laboratory, the Max-Planck-Institute for Astronomy (MPIA), the Max-Planck-Institute for Astrophysics (MPA), New Mexico State University, Ohio State University, University of Pittsburgh, University of Portsmouth, Princeton University, the United States Naval Observatory, and the University of Washington.

\bibliography{ms}

\clearpage 

\LongTables




\end{document}